\documentclass[twocolumn]{autart}    % Enable this line and disable the
\usepackage{graphicx}          % Include this line if your
\def \xs{\mathbf{x}}
\def \us{\mathbf{u}}
\def \xp{x}

\def\set[#1]{\mathcal{#1}}
\def\E{\mathrm{\textsf{E}}}
\def\nnorm[#1, #2]{\mbox{$\left. \vline\,\vline\,\displaystyle #1 \,\vline\,\vline \right._{#2}\,$}} 
\usepackage{amsmath,amsfonts,amssymb}
\usepackage{color}
\usepackage{cite}
\usepackage{graphicx}
\usepackage{algorithm}
\usepackage{algorithmic}
\usepackage{comment}
\usepackage[usenames,dvipsnames]{pstricks}
\usepackage{epsfig}
\usepackage{pst-grad} % For gradients

\def\A[#1, #2]{#1_{#2}}
\def\ti[#1, #2]{#1(#2)}
\def\Ati[#1, #2, #3]{#1_{#2}({#3})}
\def\a[#1]{{c^w_{#1}}}
\def\b[#1]{{c^d_{#1}}}
\def \x{x}

\def \bv{b}

\newtheorem{theorem}{Theorem}
\newtheorem{assumption}{Assumption}
\newtheorem{definition}{Definition}
\newtheorem{remark}{Remark}
\newtheorem{lemma}{Lemma}
\newtheorem{proposition}{Proposition} 
 
\usepackage{psfrag}

\begin{document}

\begin{frontmatter}
%\runtitle{Insert a suggested running title}  % Running title for regular
                                              % papers but only if the title
                                              % is over 5 words. Running title
                                              % is not shown in output.

\title{Constrained Distributed Algebraic Connectivity Maximization in Robotic Networks \thanksref{footnoteinfo} }  % Title, preferably not more
                                                % than 10 words.
\vskip-0.5cm\large{--- \textsc{ Technical Report } --- }

\thanks[footnoteinfo]{This paper was not presented at any IFAC
meeting. Corresponding author A.~Simonetto.}

\author[*]{Andrea Simonetto}\ead{a.simonetto@tudelft.nl},     % Add the
\author[*]{Tam\'as Keviczky}\ead{t.keviczky@tudelft.nl}, and                % e-mail address
\author[*]{Robert Babu\v{s}ka}\ead{r.babuska@tudelft.nl}

\address[*]{Delft Center for Systems and Control, Delft University of Technology, Mekelweg 2, 2628 CD Delft, The Netherlands}  % Please supply

                             % keyword list or with the
                                          % help of the Automatica
                                          % keyword wizard

\begin{abstract}
We consider the problem of maximizing the algebraic connectivity of the communication graph in a network of mobile robots by moving them into appropriate positions. We define the Laplacian of the graph as dependent on the pairwise distance between the robots and we approximate the problem as a sequence of Semi-Definite Programs (SDP). We propose a distributed solution consisting of local SDP's which use information only from nearby neighboring robots. We show that the resulting distributed optimization framework leads to feasible subproblems and through its repeated execution, the algebraic connectivity increases monotonically. Moreover, we describe how to adjust the communication load of the robots based on locally computable measures. Numerical simulations show the performance of the algorithm with respect to the centralized solution.
\end{abstract}

\end{frontmatter}

% !TEX root = Automatica2011_TReport.tex
\section{Introduction}

Teams of autonomous mobile robots that communicate with one another to achieve a common goal are considered in several applications ranging from underwater and space exploration \cite{Leonard2010, Izzo2007}, to search and  rescue \cite{Lau2008, Casper2003}, monitoring and surveillance \cite{Casbeer2005, Mathews2007}. These robots possess on-board processing capability, but the common task can only be achieved through information exchange among the members and possibly a base station. Such multi-robot teams are thus often referred to as robotic networks. Among the engineering and research questions these applications pose, maintaining connectivity between the individual robots and increasing the communication quality under given constraints, have fundamental importance. Several types of coordination and control frameworks that have been recently proposed rely on agreement protocols or consensus processes that lead to coordinated team actions \cite{Bullo2008,Keviczky2008a,Ren2008b}. Since these protocols typically assume only local communication among ``neighboring'' robots, the interconnection topology of the underlying communication graph influences greatly their effectiveness. In particular, their convergence properties are dictated by the algebraic connectivity of the communication graph \cite{Olfati-Saber2004}.

In this paper, we study distributed solutions for maximizing the algebraic connectivity of the communication graph (often denoted as $\lambda_2$) in mobile robotic networks. We note that, besides the benefit in terms of improved communication, the tools that we develop are instrumental for handling cases where network of mobile robots have other common tasks, in addition to the requirement to increase their $\lambda_2$. Examples of scenarios where our solution could, or has been used in a preliminary version, are collaborative multi-target tracking \cite{Derenick2009, Simonetto2011c} and coordination control~\cite{Derenick2010}.  For example, in \cite{Derenick2009}, the authors specifically increase the $\lambda_2$ of a special weighted graph that describes the visual connection with multiple targets. Their aim is to move a group of mobile robots in order to increase the visibility of multiple targets. In this context, the problem of $\lambda_2$ maximization could also be seen as an alternative formulation of the optimal sensing placement problem in a dynamic environment~\cite{Derenick2009}.

%We will focus on distance-based connectivity maximization with minimum separation constraints, as opposed to ensuring line-of-sight connectivity in an obstacle-rich environment \cite{anisi2008ccm}. 

References \cite{Spanos2004, Cort'es2006, Yang2008, Zavlanos2008, Zavlanos2009, Schuresko2009} give a comprehensive overview of distributed algorithms for robotic networks that aim at ensuring
connectivity (i.e., nonzero $\lambda_2$ rather than its maximization). Typically, these algorithms are either limited to specific scenarios only, or imply heavy communication requirements, and often they are not directly related to the solution of the centralized version. In terms of distributed connectivity maximization, the available literature appears to be very limited. To the best of our knowledge, only the work in \cite{DeGennaro2006} investigates a distributed solution for the maximization of $\lambda_2$ based on a simplified scenario where the dynamics of the robots are represented by a single integrator and no constraints are present. The authors use a two-step distributed algorithm, which relies on super-gradients and potential functions. The required communication load scales with the square of the graph diameter which may impede fast real-time implementations for large groups of robots.

We consider as starting point the centralized optimization procedure of \cite{Kim2006, Boyd2006, Derenick2009}. In these works the maximization of the algebraic connectivity is
approximated as a sequence of Semi-Definite Programs based on the notion of state-dependent graph Laplacian, while the agents are modeled as discrete-time single integrators.

Our first contribution is to modify the aforementioned centralized optimization procedure in order to handle more generic LTI robot dynamics. The resulting optimization problem is then proven to be feasible at each time step under quite general assumptions.

As our second contribution, we propose a distributed solution for the centralized problem (Algorithm~1) substantially extending our preliminary results in \cite{Simonetto2011}. Our proposed distributed approach relies on local problems that are solved by each robot using information only from nearby neighbors and, in contrast with \cite{DeGennaro2006}, it does not require any iterative schemes, making it more suitable for real-time applications. This last property is not a trivial aspect when using common decomposition methods for optimization \cite{Bertsekas1997}, as done in various approaches to distributed control \cite{Langbort2004a,Rantzer2009}. In our approach \emph{(i)} we formulate local problems of small size that are clearly related to the centralized one, \emph{(ii)} the \emph{linearized} algebraic connectivity of the approximate problem is guaranteed to be monotonically increasing, \emph{(iii)} the overall optimization scheme is proven to be feasible at each time step under quite general assumptions, and in particular \emph{(iv)} the local solutions are feasible with respect to the constraints of the original centralized problem. 

%Although almost all these properties are of paramount importance in real applications, they appear to be either absent or marginally considered in the available literature.

Finally, we characterize the local relative sub-optimality of the optimized $\lambda_2$ with respect to a larger neighborhood size and we use this characterization to enable each robot to increase or decrease its communication load on-line, while respecting the properties \emph{(ii)} - \emph{(iv)}. This means that our solution can be adapted based on available resources, augmenting or reducing the required communication and computational effort.

The proposed distributed solutions can be seen as a complementary approach to standard subgradient algorithms \cite{Bertsekas1997}. Distributed versions of incremental subgradient algorithms are typically communication intensive iterative algorithms, in which at each iteration, each agent has to evaluate only a local subgradient of a certain function. Our proposed solutions lie on the other side of the ``communication-computation'' trade-off spectrum. In fact, each robot solves a reasonably complex convex optimization problem, while the communication among them remains limited. In this context, multi-robot systems embedded with reasonable processing capabilities, where real-time applicability is a strong requirement, could benefit more from our proposed approach than from standard subgradient algorithms. 

%Simulation results support the efficacy of our approach and illustrate interesting properties of the algorithm with respect to centralized schemes.

%For instance, given the nonlinear/nonconvex nature of the problem, in certain scenarios the
%distributed solutions converge to a higher $\lambda_2$ value than
%the centralized ones obtained from the approximate problem
%formulation.

The paper is organized as follows. Sections~\ref{sec:prob}-\ref{sec:KIM} formulate the approximate centralized problem based on \cite{Kim2006}. Starting from a general time-invariant non-convex formulation~\eqref{eq:Kimprob}, first we discuss the sequential Semi-Definite Programming approach~\eqref{eq:centralprob} considering single integrator dynamics for the agents~\eqref{eq.dyneq}, as done in the literature~\cite{Kim2006, Derenick2009}. Second, in Section~\ref{sec:MORE} we extend this sequential Semi-Definite Programming approach to more general LTI agent dynamics~\eqref{eq:dynsyst} in problem~\eqref{eq:centralprobextended}. The proposed distributed approach for problem~\eqref{eq:centralprobextended} is described in Section~\ref{sec:distr} in problem~\eqref{eq:distributedprob} and Algorithm~1. Its properties are analyzed in Section~\ref{sec:prop}, while the local relative sub-optimality measures are the topic of Section~\ref{sec:sub}. Numerical simulations are shown in Section~\ref{sec:results} to assess the performance of the distributed solutions. Conclusions and open issues are discussed in Section~\ref{sec:future}.

\section{Problem Formulation} \label{sec:prob}

The notation is standard: for any real scalar $s$, $s \in \mathbb{R}_0$ if $s \neq 0$, $s \in \mathbb{R}^{+}$ if $s \geq 0$, and $s \in \mathbb{R}_0^+$ if $s > 0$. The matrices $I_n$ and $0_n$ represent the identity and the null matrix with dimension $n \times n$, respectively. The column vectors $\mathbf{1}_n$ and $\mathbf{0}_n$ define vectors of dimension $n$ where all the entries are $1$ and $0$, respectively.

Consider a network of $N$ agents with communication and computation capabilities and express as $a_i(k)$ the value of the variable $a$ for agent $i$ at the discrete time instant $k$. The position of agent $i$ is denoted by $x_i(k)\in \mathbb{R}^3$ and its velocity by $v_i(k)\in \mathbb{R}^3$. In order to introduce the works of \cite{Kim2006, Boyd2006, Derenick2009}, we assume the agents to move according to the following discrete-time dynamical system:
\begin{equation}
  x_i(k+1) = x_i(k) + v_i(k) T_s
  \label{eq.dyneq}
\end{equation}
where $T_s$ is the sampling time. This single-integrator model will be extended in subsequent sections. 

\begin{figure}
\footnotesize
  \setlength{\unitlength}{\textwidth}\begin{picture}(0.4,0.146)
  \put(0.085,0){\psfrag{a}{\tiny$\rho_1$}\psfrag{b}{\tiny$\rho_2$}\psfrag{c}{\hskip-1.5cm\tiny$d^2_{ij}(k) = ||\xp_i(k) - \xp_j(k)||^2$}
  \psfrag{d}{\tiny\hskip-0.2cm$f_w$}
  \includegraphics[trim=0 2.4cm 0 2cm, clip, width=0.3\textwidth]{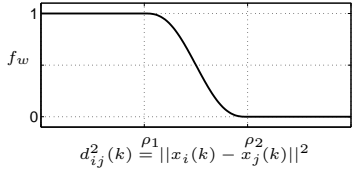}
  }
\end{picture} \\
\caption{Weighting function $f_w(\cdot)$ for modeling connectivity between two agents $i, j$. If $d^2_{ij}(k) < \rho_1$ then $w_{ij} = 1$, while if $d^2_{ij}(k) > \rho_2$ then $w_{ij} = 0$.}
\label{tab:poly}       % Give a unique label
\end{figure}

Graph-theoretic notions are used to model the network. Let $x(k)$ be the stacked vector containing the positions of the agents, i.e. $x(k)= (x_1^{\top}(k), \dots, x_N^{\top}(k))^{\top}$. The set $\mathcal{V}$ contains the indices of the mobile agents (nodes), with cardinality $N = |\mathcal{V}|$. The set $\mathcal{E}$ indicates the set of communication links. The graph $\mathcal{G}$ is then expressed as $\mathcal{G} = (\mathcal{V}, \mathcal{E})$ and it is assumed undirected. Let the agent clocks be synchronized, and assume perfect communication (no delays or packet losses). The agents with which agent $i$ communicates are called neighbors and are contained in the set $\mathcal{N}_i$. Note that agent $i$ is not included in the set $\mathcal{N}_i$. We define $\mathcal{N}^+_i = \mathcal{N}_i \cup \{i\}$ and $N_i = |\mathcal{N}^+_i|$. Define the Laplacian matrix $L$ associated with $\mathcal{G}$ via its entries $\ell_{ij}$ as $\ell_{ij}(k) = 0$ for $(i,j) \notin \mathcal{E}$, $\ell_{ij}(k) = -w_{ij}(k)$ for $(i,j) \in \mathcal{E}$, and $\ell_{ij}(k) = \sum_{l\neq i} w_{il}(k)$ for $i = j$. 
%\footnotesize\vskip-0.5cm
%\begin{equation}
%  \ell_{ij}(k) = \left\{\begin{array}{cl}
%            0    & (i,j) \notin \mathcal{E}\\
%         -w_{ij}(k) & (i,j) \in \mathcal{E}, i \neq j\\
%         \sum_{l\neq i} w_{il}(k) & i = j
%       \end{array}\right.
%       \label{eq:Laplacian}
%\end{equation}
%\normalsize
The weights $0 \leq w_{ij} \leq 1$ are assumed to depend on the squared Euclidean distance of $\xp_i(k)$ and  $\xp_j(k)$ defined as
\footnotesize\vskip-0.5cm
\begin{equation} \label{eq:dij}
  d^2_{ij}(k) = f_d(x_i(k), x_j(k)) = ||x_i(k) - x_j(k)||^2
\end{equation}
\normalsize
and
\footnotesize\vskip-0.5cm
\begin{equation} \label{eq:wij}
  w_{ij}(k) = f_{w}(||\xp_i(k) - \xp_j(k)||^2)
\end{equation}
\normalsize
where $f_{w}: \mathbb{R}^{+} \to [0,1]$ is a smooth nonlinear function with compact support. The weights model the connection strength between two agents. The closer two agents are, the closer to one is the weight, representing an increase in the communication ``quality''. For simulation purposes we use the function qualitatively represented in Figure~\ref{tab:poly}, which is one when the squared distance is less than $\rho_1$ and it is zero when the squared distance is greater than $\rho_2$. For a detailed discussion on the choice of $f_{w}$ the reader is referred to \cite{Kim2006}.
As a direct consequence of the above definitions, the entries of the Laplacian matrix $L$ depend on the state of the agents, making it state-dependent, which we will denote by $L(x(k))$.

We are interested in maximizing the algebraic connectivity of the weighted graph by controlling the state of the agents, i.e., moving them to appropriate positions. First of all, we notice that~\cite{Boyd2006}
$$
\max_{x}\lambda_2(x) \equiv \{\max_{x,\gamma} \gamma|\textrm{s.t. } L(x) + \mathbf{1}_N\mathbf{1}_N^T \succ \gamma I_N\},
$$
which can be proven formally as follows. 
\begin{proposition}\label{lemma.lmi}
For any two scalars $\lambda > \bar{\lambda}_2 > 0$, the constraint
\begin{equation}
\lambda_2(L) >  \bar{\lambda}_2,
\label{eq.barlambda2}
\end{equation}
can be formulated with the equivalent Matrix Inequality
\begin{equation}\label{eq:conlmi}
L + (\lambda/N)\mathbf{1}_N\mathbf{1}_N^{\top} \succ \bar{\lambda}_2 I_N.
\end{equation} 
\end{proposition}
\emph{Proof.} By construction, the Laplacian matrix $L$ has as eigenvector $\mathbf{e}_1 = \mathbf{1}_N$. All the other eigenvectors, $\mathbf{e}_i$, are orthogonal to $\mathbf{1}_N$, meaning $\mathbf{1}_N^\top \mathbf{e}_i = 0$, for $i = 2, \dots, N$. This implies that 
\begin{equation*}
\left(L + (\lambda/N)\mathbf{1}_N\mathbf{1}_N^{\top}\right) \mathbf{e}_i = L \mathbf{e}_i = \lambda_i \mathbf{e}_i, \quad \mathrm{for~} i = 2, \dots, N
\end{equation*}
and therefore $L + (\lambda/N)\mathbf{1}_N\mathbf{1}_N^{\top}$ has the same eigenvalues/eigenvectors of $L$ for $i = 2, \dots, N$. The remaining eigenvalue is associated with the $\mathbf{e}_1$ eigenvector:
\begin{equation*}
\left(L + (\lambda/N)\mathbf{1}_N\mathbf{1}_N^{\top}\right) \mathbf{e}_1 = L \mathbf{1}_N + (\lambda)\mathbf{1}_N = \lambda \mathbf{1}_N
\end{equation*}
and its value is $\lambda$. As a result, the eigenvalues of $L + (\lambda/N)\mathbf{1}_N\mathbf{1}_N^{\top}$ are $$
\lambda, \lambda_2(L), \lambda_3(L), \dots, \lambda_N(L).
$$ 
Since we have already that $\lambda > \bar{\lambda}_2$ (by assumption), and $\lambda_2(L) \leq \lambda_3(L) \leq \dots \lambda_N(L)$, the constraint~\eqref{eq:conlmi} imposes that $\lambda_2(L) > \bar{\lambda}_2$ and thus it is equivalent to~\eqref{eq.barlambda2}.  \hfill$\Box$

Since for the specified weighted Laplacian $L(x)$ the maximum value for $\lambda_2$ is $N-1$ \cite{deAbreu2007}, we can chose $\lambda = N$ in~\eqref{eq:conlmi} and write the maximization of $\lambda_2$ as
%\begin{subequations}
\begin{eqnarray}\label{eq:Kimprob}
 \mathbf{P}\left(L(x), \rho_1\right): \quad \displaystyle \max_{x, \gamma} && \gamma
\end{eqnarray}
%\end{subequations}
\footnotesize \vskip-1.25cm
%\begin{equation*}
\begin{eqnarray*}
 \qquad\qquad\qquad\quad \textrm{s.t.} && \gamma > 0 \\
             ~ && L(x) + \mathbf{1}_N\mathbf{1}_N^T \succ \gamma I_N\\
             ~ && f_d(x_i, x_j) > \rho_1, \quad \forall (i,j)\in \mathcal{E}
\end{eqnarray*}
%\end{equation*}
\normalsize
The optimal decision variables are the final robot locations $x$ and the optimal
value of $\gamma$ which is the maximum $\lambda_2$ for $L(x)$. The constraint on $f_d(x_i, x_j)$ prevents the agents from getting too close to each other and ensures that the trivial solution in which all the agents converge to one point is not part of the feasible solution set of~\eqref{eq:Kimprob}.

\section{Centralized Solution}\label{sec:KIM}

Problem~\eqref{eq:Kimprob} is non-convex~\cite{Kim2006} but it is rather standard to obtain a time-varying convex approximation by using first-order Taylor expansions, \cite{Kim2006, Derenick2009}. Define
\footnotesize\vskip-0.75cm
\begin{eqnarray}
  \a[{ij}] &=& \left.\frac{\partial f_{w}}{\partial d_{ij}^2}\frac{\partial d_{ij}^2}{\partial \xp_i}\right|_{\xp_i(k), \xp_j(k)} = -\left.\frac{\partial f_{w}}{\partial d_{ij}^2}\frac{\partial d_{ij}^2}{\partial \xp_j}\right|_{\xp_i(k), \xp_j(k)}, \\  \b[{ij}] &=& \left.\frac{\partial f_d}{\partial \xp_i}\right|_{\xp_i(k), \xp_j(k)} = -\left.\frac{\partial f_d}{\partial \xp_j}\right|_{\xp_i(k), \xp_j(k)}
\end{eqnarray}
\normalsize
%\begin{equation*}
%  b_{ij} = \left.\frac{\partial f_d(\xp_i, \xp_j)}{\partial \xp_i}\right|_{\xp_i(k), \xp_j(k)} %= -\left.\frac{\partial f_d(\xp_i, \xp_j)}{\partial \xp_j}\right|_{\xp_i(k), \xp_j(k)}
%\end{equation*}
then
\footnotesize\vskip-0.75cm
\begin{equation}
  w_{ij}(k+1) = w_{ij}(k) + \a[{ij}]^\top  (\delta \xp_i(k+1) - \delta \xp_j(k+1))
\label{eq.linw}
\end{equation}
\begin{equation}
  d^2_{ij}(k+1) = d^2_{ij}(k) +  \b[{ij}]^\top (\delta x_i(k+1) - \delta x_j(k+1))
  \label{eq.lind}
\end{equation}
\normalsize
where $\delta$ represents the difference operator, i.e. $\delta x_i(k+1) = x_i(k+1) - x_i(k)$. The symbol $\Delta$ will be employed to define the linearized entities; hence the entry $\Delta \ell_{ij}(x(k+1))$ of the Laplacian $\Delta L(x(k+1))$ will be
\footnotesize\vskip-0.5cm
\begin{equation}
  \Delta \ell_{ij}(x(k+1)) = \textcolor[rgb]{1.00,1.00,1.00}{[\Delta L]_{ij}(x(k+1))}
\end{equation}
\small \vskip-0.75cm
\begin{equation*}
  \left\{\begin{array}{cl}
            0    & (i,j) \notin \mathcal{E}\\
         -w_{ij}(k)- {\a[{ij}]^\top}  (\delta \xp_i(k+1) - \delta \xp_j(k+1)) & (i,j) \in \mathcal{E}, i \neq j \\
         \sum_{l\neq i} w_{il}(k+1) & i = j
       \end{array}\right.
\end{equation*}
\normalsize
while
\footnotesize\vskip-0.75cm
\begin{equation}
  \Delta f_d(\xp_i(k+1), \xp_j(k+1)) = d^2_{ij}(k) +  {\b[{ij}]^\top} (\delta x_i(k+1) - \delta x_j(k+1))
\end{equation}
\normalsize
This allows us to consider the maximization of the algebraic connectivity of $L$ as the following time-varying convex optimization problem\cite{Kim2006, Derenick2009}:
%\begin{subequations}
\begin{eqnarray}\label{eq:centralprob}
 \Delta \mathbf{P}\left(L(x(k)), x(k), \mathcal{S}_{\Delta \mathcal{Q}_2}\right): \displaystyle \max_{x(k+1), \gamma(k+1)}  \gamma(k+1)
\end{eqnarray}
\footnotesize \vskip-1.25cm
\begin{eqnarray*}
 \qquad\qquad\textrm{s.t.} && \nonumber \\
 ~ && \hskip-1.5cm \Delta \mathcal{Q}_1: \left\{ \begin{array}{c}
                                           \gamma(k+1) > 0 \\
                                           \Delta L(\x(k+1)) + \mathbf{1}_N\mathbf{1}_N^T \succ \gamma(k+1) I_N\\
                                        \end{array}\right. \\
             ~ && \hskip-1.5cm \Delta \mathcal{Q}_2: \left\{ \begin{array}{ll}
                                           \mathcal{Q}_{2.1}:& \Delta f_d(\xp_i(k+1), \xp_j(k+1))> \rho_1, \\
                                            & \qquad\quad \forall (i,j)\in \mathcal{E} \\
                                           \mathcal{Q}_{2.2}:& ||x_i(k+1) - x_i(k)|| \leq v_{\max}T_s \\ & \qquad\quad i=1,\ldots, N \\
                                        \end{array}\right.
\end{eqnarray*}
\normalsize
%\end{subequations}
where $\mathcal{S}_{\Delta \mathcal{Q}_2} = \{\rho_1, v_{\max}\}$ represents the parameter set that characterizes the set of constraints $\Delta \mathcal{Q}_2$, and it is used to highlight the dependence of the problem on the ``physical'' limitation of the application scenario (i.e., in this case, the mutual distance $\rho_1$ and the maximum allowed velocity $v_{\max}$). %It will be shown later how this parameter set will change in the different formulations of the problem. 

In contrast to the original non-convex problem~\eqref{eq:Kimprob}, the optimization problem~\eqref{eq:centralprob} is solved \emph{repeatedly} at each discrete time step $k$ on-line. In this sense~\eqref{eq:centralprob} is the $k$-th problem of a sequence of convex SDP problems. Note that the achieved maximal algebraic connectivity $\gamma$ depends on $k$ and thus we use $\gamma(k)$, while the iterative scheme for updating $\gamma$ is the repeated solution of the optimization problem itself.  This means that, letting $ \Delta \mathbf{P}(x(k))$ represent problem~~\eqref{eq:centralprob}, $\gamma$ evolves as
$$ 
(x(k+1), \gamma(k+1)) = \mathrm{arg}{\min} \,\Delta \mathbf{P}(x(k)).
$$
As a consequence of using this sequential convex programming approach (and as a consequence of the non-convex nature of the original problem), although we aim at increasing the cost function at each step $k$, we might converge to a local minimum of the original problem~\eqref{eq:Kimprob} and a strong dependence on the initial configuration of the agents has to be expected. Despite these drawbacks, it has been shown~\cite{Kim2006} that this formulation does indeed lead to satisfactory local optimal final configurations with a clear increase in the algebraic connectivity.  

Assuming that the initial positions $x(0)$ form a connected graph and the mutual distance between the agents is greater than $\sqrt{\rho_1}$, i.e., assuming initial feasibility for the problem, we can prove that the optimization problems~\eqref{eq:centralprob} will remain feasible for all the subsequent time steps $k>0$ (in fact one can always select $x(k) = x(k+1)$ to obtain a feasible solution) and their solution sequence monotonically increases the algebraic connectivity, \cite{Kim2006}. The property of remaining feasible for all $k$ is related to \emph{persistent} feasibility (also known as \emph{recursive} feasibility), which  is a well-known and fundamental concept in the optimization-based control literature~\cite{Borrelli2011}. In particular, persistent feasibility ensures that, for any $k$, if the $k$-th convex problem~\eqref{eq:centralprob} is feasible then the $(k+1)$-st problem will be feasible. This, in addition to initial feasibility (i.e., feasibility at $k=0$), guarantees that the overall sequential optimization scheme is feasible for all $k>0$. It has to be noted that persistent feasibility ensures only that the solution set of each problem~\eqref{eq:centralprob} is non-empty, while any improvement in the cost function should be proven separately. However, persistent feasibility is needed to justify the overall optimization scheme in practice. 

\section{More general LTI dynamical models}\label{sec:MORE}

As our first contribution, we extend the problem~\eqref{eq:centralprob} in order to allow a more general LTI dynamical model for the agents. Let $\xs_i(\tau) = (x_i(\tau)^{\top}, v_i(\tau)^{\top})^{\top}$ be the state of agent $i$ at the discrete time $\tau$. We note that the sampling periods belonging to $\tau$ and $k$ may differ, meaning that the optimization~\eqref{eq:centralprob} could be run at a slower rate than the system dynamics. Let the agents have the following second order discrete-time LTI dynamics:
\footnotesize\vskip-0.75cm
\begin{equation}
  \left( \begin{array}{c}
                x_i(\tau+1) \\
                v_i(\tau+1) \\
              \end{array}
            \right)
   = \left( \begin{array}{cc}
                I_3& A_{1i} \\
                0_3& A_{2i} \\
              \end{array}
            \right) \left( \begin{array}{c}
                x_i(\tau) \\
                v_i(\tau) \\
              \end{array}
            \right) + \left( \begin{array}{c}
                0_3 \\
                \bv_{1i} I_3 \\
              \end{array}
            \right) u_i(\tau)
  \label{eq:dynsyst}
\end{equation}
\normalsize
where $A_{1i} \in \mathbb{R}^{3 \times 3}$, $A_{2i}  \in \mathbb{R}^{3 \times 3}$,  $\bv_{1i}\in \mathbb{R}_0$, and $u_i(\tau)\in \mathbb{R}^3$ is the control input. Assume:

\begin{assumption}\label{assumption:fullrank}
The matrix $A_{1i}$ is full rank  $\forall i$.
\end{assumption}

\begin{assumption}\label{assumption:boundedu}
The control input for each agent at each discrete time step is constrained in the closed polytopic set $\bar{\mathcal{U}}_i$:
\footnotesize\vskip-0.75cm
\begin{equation}
  u_i(\tau) \in \bar{\mathcal{U}}_i, \, \bar{\mathcal{U}}_i = \{u_i(\tau) \in\mathbb{R}^3| H_i u_i(\tau) \leq h_i\}, \mathbf{0}_3 \in \bar{\mathcal{U}}_i
\end{equation}
\normalsize
described via the matrix $H_i$ and the vector $h_i$.  
\end{assumption}

Assumption~\ref{assumption:fullrank} is meant to ensure the one-step controllability of the dynamical system described in Eq.~\eqref{eq:dynsystk}. Analogously to $v_{\max}$ in problem~\eqref{eq:centralprob}, Assumption~\ref{assumption:boundedu} limits the control input to account for the physical limitations of the agents, and it is a standard formulation of actuator limitations in the optimization-based control community. The state space system in~\eqref{eq:dynsyst} can model agents for which the acceleration does not depend on the position and for which zero velocity and acceleration input ($v_i(\tau)=0$ and $u_i(\tau)=0$) implies $x_i(\tau+1) = x_i(\tau)$. Typically, this class of systems can represent different types of physical agents ranging from fully actuated mobile robots to underwater vehicles. The choice $A_{1i} = I_3 T_s$, $A_{2i} = I_3$, $\bv_{1i} = T_s$ yields a double integrator with sampling period $T_s$. The reason for the choice of~\eqref{eq:dynsyst} is to consider the simplest model that is capable of showing how to handle the main difficulties when extending the optimization problem~\eqref{eq:centralprob} to general LTI models. In particular, the key issues are persistent feasibility and collision avoidance. To guarantee persistent feasibility we show how to ensure that $\xs_i(k+1) = (x_i^{\top}(k), \mathbf{0}_3^{\top})^{\top}$ is a feasible state for all the agents recalling that the feasibility of the similar solution $x_i(k+1) = x_i(k)$ is a sufficient condition for~\eqref{eq:centralprob} to be persistently feasible. The collision avoidance issue is generated from the fact that the constraint on $f_d(x_i(k), x_j(k))$ is enforced only at each time step $k$, when the optimization problem is solved, but not for every $\tau$, which might be a higher rate implementation of the dynamical model. In this respect we show how to ensure that $f_d(x_i(\tau), x_j(\tau)) > 0$ for every $\tau$. We will show that when persistent feasibility and collision avoidance are handled correctly, the problem~\eqref{eq:centralprob} can be extended to dynamical models of the form~\eqref{eq:dynsyst}. In Appendix~\ref{appendix1} we discuss how to possibly cope with these two aspects for an even broader class of LTI dynamical systems.

\begin{remark}
The results of this papers apply to agents modeled via specific LTI dynamical systems. However, the paths (or waypoints) generated for these LTI agents could still be \emph{followed} by differential drive/tracked vehicles, which see widespread use in mobile robotics.  Additional examples include the works of M.~M.~Zavlanos and co-authors~(e.g., \cite{Zavlanos2011}) where the discrete-time optimization is used by a continuous-time robot (whose dynamics can be rather arbitrary) in a hybrid systems fashion.
\end{remark}

\subsection{Persistent Feasibility}
The first step to guarantee persistent feasibility is to ensure that at each time step $k$ we can affect the position of the agents via the control input. This is not trivial because the position $x_i(\tau+1)$ cannot be controlled in one step by $u_i(\tau)$. However, we can overcome this issue by solving the optimization problem at a slower rate than the implementation of the control input, e.g., once in two time steps $\tau$ when we determine both $u_i(\tau)$ and $u_i(\tau+1)$. In this case the dynamical system~\eqref{eq:dynsyst} can be lifted as seen by the optimization problem:
\footnotesize\vskip-0.85cm
\begin{multline}
  \left( \begin{array}{c}
                x_i(\tau+2) \\
                v_i(\tau+2) \\
              \end{array}
            \right)
   = \left( \begin{array}{cc}
                I_3 & A_{1i}(I_3 + A_{2i}) \\
                0_3 & A^2_{2i} \\
              \end{array}
            \right) \left( \begin{array}{c}
                x_i(\tau) \\
                v_i(\tau) \\
              \end{array}
            \right) +\\ \left( \begin{array}{cc}
                \bv_{1i}A_{1i} &  0_3 \\
                \bv_{1i}A_{2i} & \bv_{1i} I_3 \\
              \end{array}
            \right) \left( \begin{array}{c}
                u_i(\tau) \\
                u_i(\tau + 1) \\
              \end{array}
            \right)
  \label{eq:dynsystk}
\end{multline}
\normalsize
we let $k = \tau/2$, and for integer $k$'s, we define the lifted variables $x_i^L(k) = x_i(\tau)$, $v_i^L(k) = v_i(\tau)$, the lifted state $\xs^L_i(k) = (x_i^L(k)^\top, v_i^L(k)^\top)^\top$, and the lifted control input $\us^L_i(k) = (u_i(\tau)^\top, u_i(\tau+1)^\top)^\top$. For the sake of simplicity, from now on, we will omit the superscript $L$ with the idea that if we use the index $k$ we are referring to the lifted variables. With this in mind, we can rewrite the system~\eqref{eq:dynsystk} using the short-hand notation
\begin{equation}
\xs_i(k+1) = \mathcal{D}_i(\xs_i(k), \us_i(k))\label{eq:liftedsystem}
\end{equation}
We note that the lifted system~\eqref{eq:liftedsystem} is controllable to an arbitrary state in one step from $k$ to $k+1$ (see Remark~\ref{rem.2din} for details). However, the input is constrained to lie in $\us_i(k) \in \mathcal{U}_i$ (Assumption~\ref{assumption:boundedu}), where $\mathcal{U}_i = \bar{\mathcal{U}}_i\times\bar{\mathcal{U}}_i$, i.e.:
\footnotesize \vskip-.75cm
\begin{equation}
   \mathcal{U}_i = \left\{\us_i(k) \in\mathbb{R}^6\left|\left( \begin{array}{cc}
                H_i &  \\
                 & H_i \\
              \end{array}
            \right) \us_i(k) \leq \left( \begin{array}{c}
                h_i \\
               h_i \\
              \end{array}
            \right)\right. \right\}, \mathbf{0}_6 \in \mathcal{U}_i
\end{equation}
\normalsize
Therefore, the next step is to find a feasible control input value $\us_i(k) \in \mathcal{U}_i$ for which $\mathcal{D}_i(\xs_i(k), \us_i(k)) = (x_i(k)^\top, \mathbf{0}_3^\top)^\top$. For this reason define the set $\mathcal{F}_i$ as
\begin{multline}
  \xs_i(k) \in \mathcal{F}_i \Rightarrow \exists \us_i(k)\in \mathcal{U}_i\,\, \mathrm{such}\: \mathrm{that}\\  \mathcal{D}_i(\xs_i(k), \us_i(k)) = (x_i(k)^\top, \mathbf{0}_3^\top)^\top, \, \forall k \in \mathbb{N}^{+}
  \label{set.fpci}
\end{multline}
For the system~\eqref{eq:dynsystk} the set $\mathcal{F}_i$ can be computed as the Cartesian product of $\, \mathcal{F}_{x,i}$ and $\mathcal{F}_{v,i}$, i.e., $\mathcal{F}_i = \mathcal{F}_{x,i} \times \mathcal{F}_{v,i}$, where:
\vskip-0.75cm
\begin{multline*}
  \mathcal{F}_{x,i} = \left\{\x_i(k) \in\mathbb{R}^3\right\}, \qquad \textrm{  and }  \hfill\\
  \hskip10pt\mathcal{F}_{v,i} = \hfill %\\
\end{multline*}
\footnotesize \vskip-1.25cm
\begin{equation}
  \hskip-0pt\left\{v_i(k) \in\mathbb{R}^3\left| -\left( \begin{array}{c}
                H_i \bv_{1i}^{-1} (I_3 + A_{2i}) \\
                H_i \bv_{1i}^{-1}A_{2i}(I_3 + 2 A_{2i}) \\
              \end{array}
            \right) v_i(k)
   \leq \left( \begin{array}{c}
                h_i \\
               h_i \\
              \end{array}
            \right) \right. \right\}
  \label{eq:F}
\end{equation}
\normalsize
We note that $(x_i(k)^\top, \mathbf{0}_3^\top)^\top \in \mathcal{F}_i $.

\begin{figure*}
\hrule \vskip0.2cm
\begin{remark}\label{rem.2din}
The dynamical system~\eqref{eq:dynsystk}, which is the agent representation seen by the optimization problem, can be written as
\begin{equation*}
  \xs_i(k+1)
   = \left( \begin{array}{cc}
                I_3 & A_{1i}(I_3 + A_{2i}) \\
                0_3 & A^2_{2i} \\
              \end{array}
            \right) \xs_i(k)
            + \left( \begin{array}{cc}
                b_{1i}A_{1i} &  0_3 \\
                b_{1i}A_{2i} & b_{1i} I_3 \\
              \end{array}
            \right) \us_i(k) \hskip4cm (R1)
\end{equation*}
This system is controllable in one-step by an unconstrained $\mathbf{u}_i(k)$. In fact, given an arbitrary state vector $\mathbf{x}_i(k+1)$ and any initial condition $\mathbf{x}_i(k)$, due to the full rank condition on $A_{1i}$ (Assumption~1), one can promptly invert the system~(R1) and obtain the (finite) control vector $\mathbf{u}_i(k)$. To see this, consider the dynamical system~(R1) and suppose that $\mathbf{x}_i(k+1)$ and any initial condition $\mathbf{x}_i(k)$ are given. Then the control input $\mathbf{u}_i(k)$ can be determined as 
\begin{eqnarray*}
  \us_i(k)
   &=& \left( \begin{array}{cc}
                b_{1i}A_{1i} &  0_3 \\
                b_{1i}A_{2i} & b_{1i} I_3 \\
              \end{array}
            \right)^{-1}\left(\xs_i(k+1) -  \left(\begin{array}{cc}
                I_3 & A_{1i}(I_3 + A_{2i}) \\
                0_3 & A^2_{2i} \\
              \end{array}
            \right) \xs_i(k)
            \right) \nonumber \\
   &=& b_{1i}^{-1}\left( \begin{array}{cc}
                A_{1i} &  0_3 \\
                A_{2i} &  I_3 \\
              \end{array}
            \right)^{-1}\left(\xs_i(k+1) -  \left(\begin{array}{cc}
                I_3 & A_{1i}(I_3 + A_{2i}) \\
                0_3 & A^2_{2i} \\
              \end{array}
            \right) \xs_i(k)
            \right)
\end{eqnarray*}
which, is finite by Assumption~1 and $b_{1i} \in \mathbb{R}_0$.
\end{remark}
\vskip0.2cm\hrule
\end{figure*}

\subsection{Collision Avoidance}

In order to ensure no collisions, a lower bound on $\rho_1$ has to be determined, which guarantees that if $f_d(x_i(k), x_j(k)) > \rho_1$ and $f_d(x_i(k+1), x_j(k+1)) > \rho_1$, then $f_d(x_i(\tau+1), x_j(\tau+1)) > 0$, for every $k$ and $\tau$. The collision-free condition for any couple $i$ and $j$ can be written as
\footnotesize\vskip-1cm
\begin{multline}
  ||x_i(\tau+1) - x_j(\tau+1)|| \geq \\ ||x_i(\tau) - x_j(\tau)|| - ||A_{1i}v_i(\tau) - A_{1j}v_j(\tau)||  > 0
  \label{eq:condition}
\end{multline}
\normalsize
where the triangle inequality is used. Since $||x_i(\tau) - x_j(\tau)|| > \sqrt{\rho_1}$ the worst case scenario can be computed maximizing the term $||A_{1i}v_i(\tau) - A_{1j}v_j(\tau)||$ over $v_i(\tau)\in \mathcal{F}_{v, i}$  and $v_j(\tau) \in \mathcal{F}_{v, j}$. This can be rewritten as a non-convex QP problem and pre-solved off-line for any pair $i$ and $j$.\footnote{In order to see this, consider the maximizing of $||A_{1i}v_i(\tau) - A_{1j}v_{j}(\tau)||$. This is equivalent to maximize the squared norm $||A_{1i}v_i(\tau) - A_{1j}v_{j}(\tau)||^2$, which is equivalent to the following non-convex quadratic program
\begin{align*}
\max_{v_i, v_j} &&& \left(\begin{array}{c} v_i(\tau)\\ v_j(\tau)\end{array}\right)^\top \left(\begin{array}{cc} A_{1i}^\top A_{1i} & -A_{1i}^\top A_{1j} \\ -A_{1j}^\top A_{1i} & A_{1j}^\top A_{1j}\end{array}\right) \left(\begin{array}{c} v_i(\tau)\\ v_j(\tau)\end{array}\right)  \\
\textrm{subject to} &&& v_i(\tau)\in \mathcal{F}_{v, i}, \quad v_j(\tau) \in \mathcal{F}_{v, j}
\end{align*}} If $\sqrt{\bar{\rho}_1}$ denotes the worst case $||A_{1i}v_i(\tau) - A_{1j}v_j(\tau)||$ over all the pairs, then the collision-free condition~\eqref{eq:condition} can be expressed as $\rho_1  >  \bar{\rho}_1$. This is a condition that has to be imposed when designing the $\rho_1$ value in the minimal distance constraint $\mathcal{Q}_{2.1}$. In this respect, we note that the calculations performed to compute $\bar{\rho}_1$ can be made off-line before running the optimization algorithm (and therefore even the non-convex nature of the problem given the small-size and the off-line calculations can be handled in a satisfactory way in practice).

\subsection{Optimization Problem}

The optimization problem~\eqref{eq:centralprob} for the maximization of the algebraic connectivity can now be extended for the more general dynamics~\eqref{eq:dynsyst} as
\begin{equation}\label{eq:centralprobextended}
 \Delta\textbf{P} \left(\Delta L(x), \xs(k), \mathcal{S}_{\Delta \mathcal{Q}_2} \right): \quad \displaystyle \max_{\xs(k+1), \us(k), \gamma(k+1)} \gamma(k+1)
\end{equation}
\footnotesize \vskip-1.25cm
\begin{eqnarray*}
 && \hskip1.25cm \textrm{s.t.}   \\
 ~ && \hskip-0cm \Delta\mathcal{Q}_1: \left\{ \begin{array}{c}
                                           \gamma(k+1) > 0 \\
                                           \Delta L(\x(k+1)) + \mathbf{1}_N\mathbf{1}_N^T \succ \gamma(k+1) I_N\\
                                        \end{array}\right. \\
  ~ && \hskip-0cm \Delta\mathcal{Q}_2: \left\{ \begin{array}{ll}
                                           \mathcal{Q}_{2.1}:& \Delta f_d(x_i(k+1), x_j(k+1))> \rho_1, \\
                                           & \qquad \qquad \qquad \forall (i,j)\in \mathcal{E} \\
                                           \mathcal{Q}_{2.2}:& \xs_i(k+1) \in \mathcal{F}_i,  \quad i = 1, \dots, N\\
                                           \mathcal{Q}_{2.3}:& \us_i(k) \in \mathcal{U}_i,  \quad i = 1, \dots, N\\
                                           \mathcal{Q}_{2.4}:& \xs_i(k+1) = \mathcal{D}_i(\xs_i(k), \us_i(k)),\,i = 1, \dots, N\\
                                        \end{array}\right.
\end{eqnarray*}
\normalsize
where, $\mathcal{S}_{\Delta \mathcal{Q}_2} = \{\rho_1, (A_{1i}, A_{2i}, \bv_{1i}, H_i, h_i)_{i = 1, \dots, N}\}$. As a solution of~\eqref{eq:centralprobextended} we find the optimal control inputs $\us_i(k) = (u_i(\tau)^\top, u_i(\tau+1)^\top)^\top$ that drive the system~\eqref{eq:dynsyst} from $\xs_i(k)$ to $\xs_i(k+1)$. We define the concept of feasible state as follows. 

\begin{definition}\label{def.feasib}
A state $\x(k)$ is feasible if $\mathbf{x}_i(k) \in \mathcal{F}_i, \forall i $, $\Delta L(x(k)) + \mathbf{1}_N \mathbf{1}_N^\top \succ 0$, and $ d^2_{ij}(k)> \rho_1$ $\forall (i,j)\in \mathcal{E}$.
\end{definition}

For the optimization problem~\eqref{eq:centralprobextended}, as in \cite{Kim2006}, we assume initial feasibility for the first time instance:

\begin{assumption}\label{assumption:initfeasibility}
The initial state $\mathbf{x}(0)$ is a feasible state.
\end{assumption}

The following theorem states formally the persistent feasibility property:
\begin{theorem}
   If for any discrete time $k$, $\mathbf{x}(k)$ is a feasible state according to Definition~\ref{def.feasib}, then the problem~\eqref{eq:centralprobextended} will be feasible for the discrete time $k+1$.
  \label{theo:persfeas}
\end{theorem}
\emph{Proof.} Consider $\xs_i(k+1) = (x_i(k)^\top, \mathbf{0}_3^\top)^\top$  as the solution of the optimization~\eqref{eq:centralprobextended} at time $k+1$. This solution satisfies $\Delta\mathcal{Q}_1$, $\mathcal{Q}_{2.1}$, and $\mathcal{Q}_{2.2}$. Moreover, since $\xs_i(k) \in \mathcal{F}_i$ by assumption, there exist control inputs $\us_i(k) \in \mathcal{U}_i$ for all the agents for which $(x_i(k)^\top, \mathbf{0}_3^\top)^\top = \mathcal{D}_i(\xs_i(k), \us_i(k))$. Therefore the solution $\xs_i(k+1)$ satisfies $\mathcal{Q}_{2.3}$ and $\mathcal{Q}_{2.4}$ and thus the claim.  \hfill $\Box$

Combining Theorem~\ref{theo:persfeas} with Assumption~\ref{assumption:initfeasibility}, it follows that the sequence of problems~\eqref{eq:centralprobextended} is feasible for all $k > 0$. We note that persistent feasibility (Theorem~\ref{theo:persfeas}) is a fundamental property to guarantee that the overall optimization scheme remains feasible, while we show later (in the distributed case) that the sequence of solutions lead to a monotonic increase of the cost function.  

The reasons for the initial choices of $k = \tau/2$~and~$\mathcal{F}_i$ should be clearer after Theorem~\ref{theo:persfeas}. The fact that $\xs_i(k) \in \mathcal{F}_i$ guarantees that the solution $\xs_i(k+1) = (x_i(k)^\top, \mathbf{0}_3^\top)^\top$ is feasible in terms of admissible control action, which is a sufficient condition to guarantee that the optimization problem~\eqref{eq:centralprobextended} is persistently feasible.  The choice $k = \tau/2$ ensures that $\mathcal{F}_i$ is always non-empty.

%The optimization procedure~\eqref{eq:centralprobextended} described in this section solves the connectivity maximization problem in a centralized manner using linearization. In the next section, we describe an approach that allows the problem to be solved using local computation and limited communication resources, which increases the flexibility and practical applicability of the robotic network.

\section{Distributed Solution}\label{sec:distr}

In this section we present our main contribution: a non-iterative and guaranteed feasible \emph{distributed} solution to solve~\eqref{eq:centralprobextended}. We note that this is not a trivial
task, since commonly used decomposition methods for optimization problems (if applicable,
e.g.~in \cite{DeGennaro2006}) typically require iterative solutions
which may not be amenable to fast real-time implementations.

Our solution depends on subproblems which each agent solves
locally and whose size can be decided according to the available
resources. This size is influenced by the notion of an enlarged
neighborhood set, collecting all the agents whose data are available
locally at each time step $k$. The proposed distributed solution is
computed in two phases. The first step is to solve a local
optimization problem that is a small-scale \emph{modified} version of the centralized problem, in which the farthest agents (in terms of graph distance, i.e. minimum number of connecting edges) are constrained to be stationary, i.e. $\xs_i(k+1) = (x_i(k)^\top,
\mathbf{0}_3^\top)^\top$. This step is similar to a Jacobi-type
optimization \cite{Bertsekas1997}, where only certain variables are updated at a time, but also differs in the modification of the local problems and their reduced size. The second step is to share the proposed solutions within the enlarged neighborhood and combine them
using an agent-dependent positive linear combination. We note that this
sharing/combining procedure is performed just once for each
optimization step, making the overall scheme non-iterative in
contrast with commonly used consensus algorithms. The key point in the
proposed distributed solution is to \emph{jointly} construct the feasible
local problems with modified local constraints \emph{and} the positive linear combination of the solutions to
preserve feasibility of the global solution and a monotonically increasing cost function.

The most important novelty of the proposed solution is the idea of modifying the local problems and designing the merging mechanism to ensure feasibility and improvement properties for the locally merged solution with respect to the original centralized problem. This idea, as remarked in the Introduction, offers a complementary approach to standard subgradient algorithms \cite{Bertsekas1997}, which can be thought of as being on the opposite side of the ``communication-computation'' trade-off spectrum. 

\begin{figure}
\centering
 \psfrag{a}{\footnotesize$\mathcal{N}^+_i$} \psfrag{b}{\footnotesize$i$}
 \psfrag{c}{\footnotesize$\mathcal{J}_i$} \psfrag{d}{\footnotesize$\partial\mathcal{J}_i$}
 \psfrag{i}{\footnotesize$ $}\psfrag{j}{\footnotesize$ $}
 \psfrag{r}{\footnotesize$ $}
    \includegraphics[width=0.275\textwidth]{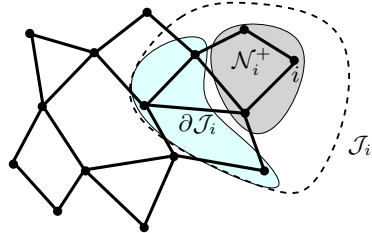}
    \caption{Notation for the distributed solution in case the size of the enlarged neighborhood size for agent $i$ is $n_i = 2$. The thick lines represent links between connected agents.}
    \label{fig.graph}
\end{figure}

Let $\mathcal{J}_i$ denote the enlarged neighborhood of $i$ consisting
of all the agents whose state is known by agent $i$ at each sampling time $k$ (either through direct or indirect communication). We define this set in a
recursive way: let $\mathcal{N}^{1}_i$ be the standard, first-order neighborhood of $i$, i.e. $\mathcal{N}^{1}_i = \mathcal{N}^+_i$, then, the $n_i$-size enlarged neighborhood of $i$ for $n_i > 1$ is defined as
\footnotesize\vskip-0.5cm
\begin{equation}
\mathcal{J}_i = \mathcal{N}^{n_i}_i = \bigcup_{j \in \mathcal{N}^{n_i-1}_i} \mathcal{N}^{n_i-1}_j
\end{equation}
\normalsize
in other words, the collection of the $(n_i-1)$-size enlarged neighborhoods of all $j \in \mathcal{N}^{n_i-1}_i$. The scalar $n_i \geq 1$ implies bounds on the diameter of the communication graph constructed with the agents in $\mathcal{J}_i$. The cardinality of $\mathcal{J}_i$ is $J_i$. We call the set of agents belonging to $\partial
\mathcal{J}_i$, the bordering agents of $\mathcal{J}_i$ defined as
\footnotesize\vskip-0.5cm
\begin{equation}
\partial \mathcal{J}_i = \{ j | j \in \mathcal{J}_i, j \notin \mathcal{N}^{n_i-1}_i \}
\end{equation}
\normalsize
Denote the graph Laplacian associated with the communication graph corresponding to the agents in $\mathcal{J}_i$ as $L_{i, n_i}$ and the communication link set as $\mathcal{E}_{i, n_i}$. Figure~\ref{fig.graph} provides a graphical illustration of this notation for $n_i = 2$.
Define $\mathbf{x}_{\mathcal{J}_i}$ and $\mathbf{u}_{\mathcal{J}_i}$ as the stacked vectors collecting the states and the lifted control inputs for all the agents $j$ belonging to the enlarged neighborhood of $i$, i.e. $j \in \mathcal{J}_i$.

As a first step of the distributed solution, for each agent $i$, we consider local problems $\Delta \textbf{P}_{i}$ of the form:
\begin{eqnarray}\label{eq:distributedprob}
\Delta \textbf{P}_{i}(\Delta L_{i, n_i}(x_{\mathcal{J}_i}), \xs_{\mathcal{J}_i}(k), \mathcal{S}_{\Delta \hat{\mathcal{Q}}_{2i}}) : && \\ \qquad \displaystyle \max_{\xs_{\mathcal{J}_i}(k+1), \us_{\mathcal{J}_i}(k), \gamma_i(k+1)} && \gamma_i(k+1)  \nonumber
\end{eqnarray}
\footnotesize\vskip-1.25cm
\begin{eqnarray*}
 && \hskip1.25cm \textrm{s.t.}   \\
 ~ && \hskip-0cm \Delta\mathcal{Q}_1: \left\{ \begin{array}{c}
                                           \gamma_{i}(k+1) > 0 \\
                                           \Delta L_{i, n_i}(\x_{\mathcal{J}_i}(k+1)) + \mathbf{1}_{J_i}\mathbf{1}_{J_i}^T \succ \gamma_{i}(k+1) I_{J_i}\\
                                        \end{array}\right. \\
  ~ && \hskip-0cm \Delta\hat{\mathcal{Q}}_{2i}: \left\{ \begin{array}{ll}
                                           \hat{\mathcal{Q}}_{2.1}:& \Delta f_d(x_i(k+1), x_j(k+1))> \hat{\rho}_{1ij},\\&  \qquad \qquad \qquad \forall (i,j)\in \mathcal{E}_{i, n_i} \\
                                           \hat{\mathcal{Q}}_{2.2}:& \xs_j(k+1) \in \hat{\mathcal{F}}_j = \mathcal{F}_j,  \quad j\in\mathcal{J}_i\\
                                           \hat{\mathcal{Q}}_{2.3}:& \us_j(k) \in \hat{\mathcal{U}}_j,  \quad j\in\mathcal{J}_i\\
                                           \hat{\mathcal{Q}}_{2.4}:& \xs_j(k+1) = \hat{\mathcal{D}}_j(\xs_j(k), \us_j(k)),\,j\in\mathcal{J}_i\\
                                        \end{array}\right. \\
  ~ && \hskip-0cm {\mathcal{Q}}_{3}: \xs_j(k+1) = (x_j(k)^\top, \mathbf{0}_3^\top)^\top, \qquad \textrm{for } j \in \partial \mathcal{J}_i
\end{eqnarray*}
\normalsize
Where $\mathcal{S}_{\Delta\hat{\mathcal{Q}}_{2i}} = \{(\hat{\rho}_{1ij}, \hat{A}_{1j}, \hat{A}_{2j}, \hat{\bv}_{1j}, \hat{H}_j, \hat{h}_j)_{j \in \mathcal{J}_i}\}$ and the notation $\hat{\mathcal{D}}_j$ denotes a dynamical system of the same form as~\eqref{eq:dynsystk} but with the modified triplet $(\hat{A}_{1j}, \hat{A}_{2j}, \hat{\bv}_{1j})$. We will show later (Theorem~\ref{theo:Q2}) how to construct the modified state matrices and parameter set $\mathcal{S}_{\Delta\hat{\mathcal{Q}}_{2i}}$.% of constraint ${\Delta\hat{\mathcal{Q}}_{2i}}$. 

The optimal local decision variables (solution of $\Delta \textbf{P}_{i}$) will be denoted as $\hat{\gamma}_{i}(k+1)$, $\hat{\xs}_{\mathcal{J}_i}(k+1)$, and $\hat{\us}_{\mathcal{J}_i}(k)$ respectively. We call $\hat{\xs}_{ij}(k+1)$ the state of agent $j$ as computed by agent $i$ and we use the same notation for $\hat{\us}_{ij}(k)$. We note that the optimal local decision variables $\hat{\xs}_{\mathcal{J}_i}(k+1)$ and $\hat{\us}_{\mathcal{J}_i}(k)$ are composed of $\hat{\xs}_{ij}(k+1)$ and $\hat{\us}_{ij}(k)$ for each $j \in
\mathcal{J}_i$. We emphasize that the extra constraint $\mathcal{Q}_{3}$ is an important requirement to guarantee feasibility, as will be explained shortly in this section. We will also require $\hat{\mathcal{F}}_i = \mathcal{F}_i$ for all the agents as a sufficient condition of persistent feasibility.

Consider the set of all agents $p$ which have agent $i$ inside their local problems $\Delta \textbf{P}_p$, i.e. $i \in \mathcal{J}_p$, and denote by $\mathcal{J}_i^* = \{ p | i \in \mathcal{J}_p \}$. Since the enlarged neighborhood size $n_i$ could differ from agent to agent, $\mathcal{J}_i^* \neq \mathcal{J}_i$.

As a second step of the distributed solution, we construct the position update based on the previous solution $\mathbf{x}(k)$ and a positive linear combination of the local position solutions $\hat{x}_{\mathcal{J}_i}(k)$ as:
\begin{equation}
 x_i(k+1) =  x_i(k) + \sum_{j \in  \mathcal{J}_i^*} \alpha_j \delta \hat{x}_{ji}(k+1), \qquad i=1, \ldots, N
  \label{eq:mean}
\end{equation}
for $\alpha_j > 0$ (recall that, since $\hat{x}_{ij}(k) = x_i(k)$, $\delta \hat{x}_{ji}(k+1) =  \hat{x}_{ji}(k+1) - {x}_{i}(k)$). Define
\footnotesize \vskip-.5cm
\begin{equation*}
\bar{\alpha}_i = \sum_{j\in\mathcal{J}_i^*} \alpha_j
\end{equation*}\normalsize
and observe that $\bar{\alpha}_i$ is in general not equal to one. We require $\bar{\alpha}_i \leq 1$ due to the linearization procedure, in fact, bigger $\bar{\alpha}_i$ would question the validity of the Taylor expansions in the local problems. 

We prove the following lemma regarding the sum of local position solutions, which is instrumental for the subsequent theorems.
\begin{lemma}
For arbitrary vectors $q_{ij} \in \mathbb{R}^3$ where $(i, j)$ are neighbors (i.e. if $\ell_{ij} \neq 0$), and for any $\delta \hat{x}_{pi}(k+1), \delta \hat{x}_{pj}(k+1)$ part of the optimal solutions of the local problems $\Delta \mathbf{P}_p$ in~\eqref{eq:distributedprob}, with $p \in \mathcal{J}_i^*$ and $p \in \mathcal{J}_j^*$ respectively, the following equality holds:
\begin{multline}
    q_{ij}^\top \left(\sum_{p \in  \mathcal{J}_i^*} \alpha_p \delta \hat{x}_{pi}(k+1) - \sum_{p \in  \mathcal{J}_j^*} \alpha_p \delta \hat{x}_{pj}(k+1) \right) = \\
    q_{ij}^\top  \sum_{p \in  \mathcal{J}_i^*\cap \mathcal{J}_j^*} (\delta \hat{x}_{pi}(k+1) - \delta \hat{x}_{pj}(k+1))
    \label{eq:dummy21}
\end{multline}
  \label{lemma.deltaax}
\end{lemma}
\emph{Proof.} The first term of the equality~\eqref{eq:dummy21} can be divided into three parts: $p \in
\mathcal{J}_i^*\cap \mathcal{J}_j^*$, $p \in \mathcal{J}_i^* \wedge p
\notin \mathcal{J}_j^*$, and $p \in \mathcal{J}_j^* \wedge p \notin
\mathcal{J}_i^*$. Since we are interested in the case when $i$ and $j$ are neighbors, we can make the key observations that:
\footnotesize \vskip-.5cm
\begin{equation}
  \{p| p \in \mathcal{J}_i^* \wedge p \notin \mathcal{J}_j^*\} \Rightarrow i \in \partial \mathcal{J}_p
  \label{eq:impl1}
\end{equation}
\begin{equation}
  \{p| p \in \mathcal{J}_j^* \wedge p \notin \mathcal{J}_i^*\} \Rightarrow j \in \partial \mathcal{J}_p
  \label{eq:impl2}
\end{equation}
\normalsize
Consider the first implication~\eqref{eq:impl1}. If $p \in \mathcal{J}_i^*$, then $i$ and $p$ are separated by at most $n_p$ links. Furthermore, if $p \notin \mathcal{J}_j^*$, then $j$ and $p$ are separated by at least $n_p+1$ links. Since $i$ and $j$ are neighbors, it follows that the separation between $i$ and $p$ is exactly $n_p$ links and therefore $i \in \partial \mathcal{J}_p$. The second implication~\eqref{eq:impl2} can be proven by similar arguments. The two implications~\eqref{eq:impl1}-\eqref{eq:impl2} allow us to rewrite the first part of the equality~\eqref{eq:dummy21} as:
\footnotesize\vskip-0.75cm
\begin{multline*}
    q^\top_{ij} \sum_{p \in \mathcal{J}_i^*\cap \mathcal{J}_j^*} \alpha_p(\delta \hat{x}_{pi}(k+1) - \delta \hat{x}_{pj}(k+1)) +\\ q^\top_{ij}\underbrace{\sum_{p \in \mathcal{J}_i^* \wedge p \notin \mathcal{J}_j^*}\alpha_p\delta \hat{x}_{pi}(k+1)}_{= 0} - q^\top_{ij}\underbrace{\sum_{p \in \mathcal{J}_j^* \wedge p \notin \mathcal{J}_i^*} \alpha_p\delta \hat{x}_{pj}(k+1)}_{= 0}
  \end{multline*}
\normalsize where the last two terms are $0$ due to~\eqref{eq:impl1}-\eqref{eq:impl2} and the constraint $\mathcal{Q}_3$ of $\Delta \mathbf{P}_p$ in~\eqref{eq:distributedprob}, which requires $\delta \hat{x}_{pi}(k+1) = 0$ and $\delta \hat{x}_{pj}(k+1) = 0$ for $i \in \partial \mathcal{J}_p$ and $j \in \partial \mathcal{J}_p$, respectively.\hfill $\Box$

We are ready to construct the parameter set $\mathcal{S}_{\Delta \hat{\mathcal{Q}}_{2i}}$ which defines the local set of constraints $\Delta \hat{\mathcal{Q}}_{2i}$.

\begin{theorem} \emph{(Local constraints for global feasibility)}
Taking for each $i$, the following choices:
\begin{itemize}
  \item the local parameter set $\Delta \hat{\mathcal{Q}}_{2i}$ in~\eqref{eq:distributedprob} as
  \begin{equation*}
\mathcal{S}_{\Delta\hat{\mathcal{Q}}_{2i}} = \{(\hat{\rho}_{1ij}, \bar{\alpha}_j^{-1}{A}_{1j}, {A}_{2j}, \bar{\alpha}_j{\bv}_{1j}, {H}_j, \bar{\alpha}_j^{-1}{h}_j)_{j \in \mathcal{J}_i}\}
\end{equation*}
meaning: $\tilde{A}_{1j} = \bar{\alpha}_j^{-1}{A}_{1j}, \tilde{A}_{2j} = {A}_{2j},  \tilde{\bv}_{1j} = \bar{\alpha}_j{\bv}_{1j}$, $\tilde{H}_j~=~{H}_j, \tilde{h}_j = \bar{\alpha}_j^{-1}{h}_j$, and
\begin{equation}
  \hat{\rho}_{1ij} = \bar{\alpha}_{ij}^{-1}\left(\rho_1 + d^2_{ij}(k)\left(\bar{\alpha}_{ij} -1\right) \right)
\label{eq:rhotilde}
\end{equation}
with $\bar{\alpha}_{ij} = \sum_{p \in
\mathcal{J}_i^*\cap \mathcal{J}_j^*} \alpha_p$;
    \item the positive linear combination of the local optimal control inputs $\hat{\mathbf{u}}_{ji}(k)$ in~\eqref{eq:distributedprob} as
        \begin{equation}
  \mathbf{u}_i(k) = \sum_{j \in  \mathcal{J}_i^*} \alpha_j  \hat{\mathbf{u}}_{ji}(k)
  \label{eq:umean}
\end{equation}
\item  the positive linear combination of the local optimal velocities $\hat{v}_{ji}(k+1)$ in~\eqref{eq:distributedprob} as
\begin{equation}
  \quad v_i(k+1) = \frac{\sum_{j \in  \mathcal{J}_i^*} \alpha_j  \hat{v}_{ji}(k+1)}{\bar{\alpha}_i}
  \label{eq:vmean}
\end{equation}
\end{itemize}
ensure that the updated position vector $\xp(k+1)$, the control vector $\mathbf{u}(k)$, and velocity vector ${v}(k+1)$ based on~\eqref{eq:mean}, \eqref{eq:umean}, and \eqref{eq:vmean} respectively, satisfy the set of constraints $\Delta \mathcal{Q}_{2}$ of the global problem~\eqref{eq:centralprobextended}.
\label{theo:Q2}
\end{theorem}
\emph{Proof.} We give a constructive proof of the theorem in Appendix~\ref{appendix.c}.

Theorem~\ref{theo:Q2} not only gives a procedure to construct the local constraints so that the linear combination~\eqref{eq:mean} satisfies the global constraints, it also establishes a link between the local quantities and the global ones. Furthermore, it ensures that in order to move to the updated state $\xs_i(k+1)$ (comprised of the position update $x_i(k+1)$ and the velocity update $v_i(k+1)$) each agent can implement the linear combination of the lifted control input~\eqref{eq:umean} as summarized in Algorithm~1, without explicitly computing the merging~\eqref{eq:mean} or \eqref{eq:vmean}. In this context, from a control perspective, each local optimization problem~\eqref{eq:distributedprob} computes the control input $\hat{\us}_{\mathcal{J}_i}(k)$. This is merged via~\eqref{eq:umean} with the enlarged neighborhood information and then implemented. This merged control by Theorem~\ref{theo:Q2} induces the merging mechanisms on the state~\eqref{eq:mean} and \eqref{eq:vmean}, which therefore do not have to be computed explicitly.

\begin{algorithm}
\caption{Distributed $\lambda_2$ Maximization.}
\label{algDSPD}
\footnotesize
%\begin{algorithmic}[1]
 %\For each agent $i$:
  %\STATE 
  1. Input for each agent $i$: $\xs_j(k), \, j \in \mathcal{J}_i$\\
  %\STATE 
  2. Solve: $\Delta \mathbf{P}_{i} \textrm{ in~(23) computing }$ 
  $$
  (\hat{\xs}_{ji}(k+1), \hat{\us}_{ji}(k+1)), \, j \in \mathcal{J}_i
  $$
  %\STATE 
  3. Communicate: $\hat{\us}_{ji}(k+1)$ among members of $\mathcal{J}_i$\\
  %\STATE 
  4. Positive linear combination:
  \begin{equation*}
    \us_i(k) = \sum_{j \in  \mathcal{J}_i^*} \alpha_j  \hat{\us}_{ji}(k)
  \end{equation*}
  %\STATE 
  5. Implement the control action $\us_i(k)$
 %\ENDFOR
%  \end{algorithmic}
\end{algorithm}
\normalsize

\section{Properties of the Distributed Solution}\label{sec:prop}

In the previous section we have seen how to construct the local problem parameter set $\mathcal{S}_{\Delta\hat{\mathcal{Q}}_{2i}}$ and positive linear combinations of the local solutions to ensure that the combined solution $(\mathbf{x}(k+1), \us(k))$ satisfies the constraint $\Delta \mathcal{Q}_2$ of the global problem~\eqref{eq:centralprob}. In this section we will look at $\Delta \mathcal{Q}_1$ and at the persistent feasibility of Algorithm~1. Theorem~\ref{theo:c10} and~\ref{theo:L} will establish that %\vskip0.1cm
\begin{itemize}
\item[] \vskip-0.2cm\textbf{(C1)} The algebraic connectivity of the global linearized
Laplacian $\Delta L(x(k+1))$ of~~\eqref{eq:centralprob} with $x(k+1)$ computed via~\eqref{eq:mean} is monotonically increasing in each iteration, which implies that $x(k+1)$ will also satisfy $\Delta \mathcal{Q}_1$ of the global problem~\eqref{eq:centralprob} for a certain value of $\gamma(k+1)\geq \gamma(k)$.
\end{itemize}
\vskip-0.2cm This is proven linking  the linear combination~\eqref{eq:mean} and the algebraic connectivity through the linear dependence of the linearized Laplacian on the position $x$. Theorem~\ref{theo:DPF} will show that  
\begin{itemize}
\item[] \vskip-0.2cm\textbf{(C2)} The distributed optimization in Algorithm~1 is persistently feasible using the constructed $\Delta \hat{\mathcal{Q}}_{2i}$'s in Theorem~\ref{theo:Q2}.
\end{itemize}
\vskip-0.2cm This is proven by the use of the relation between local and global feasibility of Theorem~\ref{theo:Q2}.

First of all, reconsider the linearized Laplacian $\triangle L (x(k+1))$ entries, given in \eqref{eq.deltalij}. We can rewrite $\triangle L (x(k+1))$ as a sum
\begin{equation*}
  \triangle L (x(k+1)) = \triangle L (\delta x(k+1)) + L (x(k)).
%\label{eq.todo}
\end{equation*}
Under the validity of the employed Taylor approximation, we assume that for all practical situations the value of $L (x(k))$ is equivalent to its linearized approximation $\triangle L (x(k))$, and therefore we can write
\begin{equation}
  \triangle L (x(k+1)) = \triangle L (\delta x(k+1)) + \triangle L (x(k)) = L(x(k+1)).
\label{eq.todo}
\end{equation}

Consider the local problem $\Delta \mathbf{P}_{i}$ in~\eqref{eq:distributedprob}, and its solution comprised of $\hat{x}_{ij}(k+1)$ for all $j \in \mathcal{J}_i$. Construct the global vector $\hat{x}^{(i)}(k+1)$  whose entries are determined based on the local solution as 
\begin{multline}
  \hat{x}^{(i)}(k+1)= (\dots, \hat{x}^{(i)}_j(k+1)^\top, \dots)^\top, \quad j = 1, \dots, N \\  \textrm{with }\hat{x}^{(i)}_j(k+1)= \left\{\begin{array}{lr}
   \hat{x}_{ij}(k+1) & \textrm{ if } j \in \mathcal{J}_i \\
    x_j(k) & \textrm{ otherwise }
  \end{array}\right.
  \label{eq:localsolution}
\end{multline}
where we keep those agent positions that have not been optimized fixed, and we update the rest from the solution of the local problem.

\begin{theorem}\emph{(C1.a)}
The positions $\hat{x}^{(i)}(k+1)$ in~\eqref{eq:localsolution} constructed from the solution of the local problem $\Delta \mathbf{P}_{i}$ in~\eqref{eq:distributedprob}, monotonically increase the algebraic connectivity of the Laplacian matrix: 
\begin{equation}
\triangle L(\tilde{x}^{(i)}(k+1)) \succeq \triangle L(x(k)).\label{eq.rel013by}\end{equation} \label{theo:c10}
\end{theorem}

\emph{Proof.} Since $\triangle L$ depends linearly on the position $x$ by~\eqref{eq.todo} we can write
$$
\triangle L(\tilde{x}^{(i)}(k+1)) =  \triangle L(\delta \tilde{x}^{(i)}(k+1)) + \triangle L(x(k)),
$$
thus the relation~\eqref{eq.rel013by} can be interpreted as
\begin{equation}
\triangle L(\delta \tilde{x}^{(i)}(k+1)) \succeq 0.
\label{eq.rel013by2}
\end{equation}

We recall that, 

\emph{First:} for~\eqref{eq:localsolution} $\delta \tilde{x}^{(i)}_j(k+1) = 0$ if $j \notin \mathcal{J}_i$. 

\emph{Second:} for the constraint $\mathcal{Q}_3$ in the local problem $\triangle \mathbf{P}_{i}$~\eqref{eq:distributedprob}, $\delta \tilde{x}^{(i)}_j(k+1) = 0$ if $j \in \partial\mathcal{J}_i$. 

For these two observations, $[\triangle L(\delta \tilde{x}^{(i)}(k+1))]_{ij} \neq 0$ only if $(i,j) \in \mathcal{E}_{i, n_i}$ and therefore up to a reodering the Laplacian $\triangle L(\delta \tilde{x}^{(i)}(k+1))$ has the form
\begin{equation}
  \left[%
\displaystyle\begin{array}{c|c}
  \displaystyle\triangle L_{i, n_i}(\delta \tilde{x}_{\mathcal{J}_i}(k+1)) & 0  \\ \hline 
  0 & 0 \\
\end{array}%
\right] \succeq 0. \label{eq.claim}
\end{equation}
We recall that $\tilde{x}_{\mathcal{J}_i}(k+1)$ is the optimal decision variable for the position in the local optimization problems (and the order of the single elements is not important). 

We can now restate~\eqref{eq.rel013by2} via~\eqref{eq.claim} as 
$$
\triangle L_{i, n_i}(\delta \tilde{x}_{\mathcal{J}_i}(k+1))  \succeq 0
$$
or 
$$
\triangle L_{i, n_i}(\tilde{x}_{\mathcal{J}_i}(k+1))  \succeq \triangle L_{i, n_i}(\tilde{x}_{\mathcal{J}_i}(k))
$$
which is true due to the local optimality of the local solution of $\triangle \mathbf{P}_{i}$.\hfill $\Box$

We can relate the positions $\hat{x}^{(i)}(k+1)$ in~\eqref{eq:localsolution} with $x_i(k+1)$ in~\eqref{eq:mean}, by the following Lemma.
\begin{lemma}
 When considering the positions $\hat{x}^{(i)}(k+1)$ in~\eqref{eq:localsolution} and $x_i(k+1)$ in~\eqref{eq:mean} the following equality holds:
 \begin{equation}
\label{eq:delta}
  \Delta L(\delta x(k+1)) = \sum_{i=1}^N \alpha_i \Delta L(\delta \hat{x}^{(i)}(k+1))
\end{equation}
  \label{lemma.deltax}
\end{lemma}
\emph{Proof.} Let us consider the entry $(i, j)$ of the Laplacian
$\Delta L$ on both sides of the expression. For the right side, $\ell_{ij}^{\mathrm{right}}$
can be expressed as
\footnotesize\vskip-.5cm
\begin{equation*}
  \ell_{ij}^{\mathrm{right}} = \a[ij]^\top \sum_{p \in \mathcal{J}_i^*\cap \mathcal{J}_j^*} \alpha_p (\delta \hat{x}_{pi}(k+1) - \delta \hat{x}_{pj}(k+1))
\end{equation*}\normalsize
since the entry $(i,j)$ will exist only for the subproblems $\Delta \mathbf{P}_p$ with $p \in \mathcal{J}_i^*\cap \mathcal{J}_j^*$. For the left side,
\footnotesize\vskip-.75cm
\begin{multline*}
    \ell_{ij}^{\mathrm{left}} = \a[ij]^\top \left(\delta x_{i}(k+1) - \delta x_{j}(k+1) \right) =\\
    \a[ij]^\top \left(\sum_{p \in  \mathcal{J}_i^*} \alpha_p \delta \hat{x}_{pi}(k+1) - \sum_{p \in  \mathcal{J}_j^*} \alpha_p \delta \hat{x}_{pj}(k+1) \right)
\end{multline*}\normalsize
The coefficient $\a[ij]^\top$ is non-zero only if $(i, j)$ are neighbors and using Lemma~\ref{lemma.deltaax}  leads to
\footnotesize\vskip-.5cm
\begin{equation*}
    \ell_{ij}^{\mathrm{left}} = \a[ij]^\top \sum_{p \in \mathcal{J}_i^*\cap \mathcal{J}_j^*} \alpha_p(\delta \hat{x}_{pi}(k+1) - \delta \hat{x}_{pj}(k+1))\qquad\hfill \Box
  \end{equation*}\normalsize
%~\hfill $\Box$

Using Theorem~\ref{theo:c10} and Lemma~\ref{lemma.deltax} we can now prove the monotonically increasing property of the algebraic connectivity of the global linearized Laplacian $\Delta L(x(k+1))$.
\begin{theorem} \emph{(C1.b)}
The algebraic connectivity of the global linearized Laplacian $\Delta L(x(k+1))$ is monotonically increasing in each iteration, meaning $\Delta L(x(k+1)) \succeq \Delta L(x(k))$, where $x(k+1)$ is computed by the combination~\eqref{eq:mean}. \label{theo:L}
\end{theorem}
\emph{Proof.} Theorem~\ref{theo:c10} implies $\triangle L(\delta \tilde{x}^{(i)}(k+1)) \succeq 0$ for all $i$. Thus summing over all agents leads to
\begin{equation*}
  \sum_{i=1}^N \alpha_i \triangle L(\delta \tilde{x}^{(i)}(k+1)) \succeq 0
\end{equation*}
\normalsize
Considering the linear combination $x_i(k+1)$ in~\eqref{eq:mean}, and the associated global vector $x(k+1)$, by Lemma~\ref{lemma.deltax} it follows that $\triangle L(\delta x(k+1)) \succeq 0$. From the linear dependence of $\triangle L$ on $x$ (Equation~\eqref{eq.todo}),
$$
\triangle L(x(k+1)) = \triangle L(\delta x(k+1)) + \triangle L(x(k))
$$
and therefore it follows that $\triangle L( x(k+1)) - \triangle L(x(k)) \succeq 0$ and the desired property: $\triangle L(x(k+1)) \succeq \triangle L(x(k))$.  \hfill$\Box$

Finally, we can show the persistent feasibility of the distributed optimization
algorithm (Algorithm~1).
\begin{theorem}\emph{(C2)} The distributed optimization
algorithm presented in Algorithm~1 is persistently feasible.
\label{theo:DPF}
\end{theorem}
\emph{Proof.}
We have to prove that if, for any discrete time $k$, $\xs(k)$ is a feasible initial state for the global optimization problem $\Delta \textbf{P}$~\eqref{eq:centralprobextended} at the discrete time $k$ (Definition~\ref{def.feasib}), then there will be a feasible solution to the distributed optimization problem in Algorithm~\ref{algDSPD}. Such a feasible solution can be thought of as an initial state $\xs(k+1)$ for the global optimization problem $\Delta \textbf{P}$~\eqref{eq:centralprobextended} at the discrete time $k+1$. We prove the existence of such feasible solution in two steps. \\
\emph{Step 1.} Using the assumption that $\xs(k)$ is a feasible initial state for the global optimization problem $\Delta \textbf{P}$~\eqref{eq:centralprobextended} at time step $k$, we can show that $\xs(k)$ is also a feasible initial state for the local problems $\Delta \mathbf{P}_i$~\eqref{eq:distributedprob}, which therefore are feasible and deliver local solutions $(\hat{\xs}_{\mathcal{J}_i}(k+1), \hat{\us}_{\mathcal{J}_i}(k+1))$ satisfying the constraints $\Delta \mathcal{Q}_1$, $\Delta \tilde{\mathcal{Q}}_{2i}$, and $\mathcal{Q}_3$. This claim follows from Theorem~\ref{theo:Q2}, in particular from the fact that $\hat{\rho}_{1ij} \leq \rho_1$. In fact, from the assumption $\bar{\alpha} \leq 1$ and $d^2_{ij}(k) > \rho_1$ (feasibility at $k$), the relation~\eqref{eq:rhotilde} yields $\hat{\rho}_{1ij} \leq \rho_1$, and thus $\xs(k)$ is also a feasible initial state for the local problems $\Delta \mathbf{P}_i$~\eqref{eq:distributedprob}.\\
\emph{Step 2.} We can show that after merging/combining the resulting local solutions $(\hat{\xs}_{\mathcal{J}_i}(k+1), \hat{\us}_{\mathcal{J}_i}(k+1))$, the final distributed state solution~$\xs(k+1)$ will be a feasible initial state for the global optimization problem $\Delta \textbf{P}$ in~\eqref{eq:centralprobextended} at the discrete time $k+1$. This second step follows directly from Theorem~\ref{theo:Q2} and Theorem~\ref{theo:L}.\hfill $\Box$ 

Similarly to Theorems~\ref{theo:Q2} and~\ref{theo:L}, we note that Theorem~\ref{theo:DPF} holds also if the agents change the size of their enlarged neighborhood $n_i$ from time step $k$ to $k+1$, since the feasibility of the state in the local problems does not depend on the enlarged neighborhood size of $\mathcal{J}_i$. This fact will be used in the next section to allow adjusting the communication load of each agent and make Algorithm~1 adaptive.

\section{Adapting the Communication Load}\label{sec:sub}

In this section we investigate further the properties of the distributed solution presented in Section~\ref{sec:distr}. First we show in Theorem~\ref{theo:Equivalence} that if all-to-all communication is allowed then the distributed solution of Algorithm~1 is equivalent\footnote{Meaning that the two solutions (centralized and distributed) are the same.} to the centralized approach in~\eqref{eq:centralprobextended}. Then we prove in Theorem~\ref{theo:nton1} that starting from the same state vector $\xs(k)$, if we run Algorithm~1 with different enlarged neighborhood sizes, the solution that delivers a higher algebraic connectivity at time step $k+1$ is the one with the greater neighborhood size $n$. This last fact enables us to characterize a local relative sub-optimality measure with respect to a larger enlarged neighborhood size.

\begin{theorem} \emph{(Equivalence)}
The distributed solution of Algorithm~1 is equivalent to the centralized one of~\eqref{eq:centralprobextended}, if all-to-all communication is allowed (meaning $n_i = N$, $\forall i$, and thus no bordering
agents) and if $\alpha_i = 1/N$, $\forall i$, is chosen as weight in the positive linear combinations of the local states and inputs~\eqref{eq:mean}, \eqref{eq:vmean}, and~\eqref{eq:umean}.
\label{theo:Equivalence}
\end{theorem}
\emph{Proof.} Consider $n_i = N$, $\partial \mathcal{J}_i =
\{\emptyset\}$ for all the agents, and the choice $\alpha_i = 1/N$,
$\forall i$. We have $\bar{\alpha}_i = 1$ and $\sum_{p \in
\mathcal{J}_i^* \cap \mathcal{J}_i^*} \alpha_p = 1$. Therefore, as a
consequence of the choices of Theorem~\ref{theo:Q2}, $\Delta \hat{\mathcal{Q}}_{2i} \equiv \Delta
{\mathcal{Q}}_{2}$. Furthermore, all the constructed local solutions $\hat{x}^{(i)}(k+1)$ in~\eqref{eq:localsolution} are the same and they are equivalent to the solution of the centralized problem $x(k+1)$ in~\eqref{eq:centralprobextended}. Given the specified selection of $\alpha_i$, also the linear combination~\eqref{eq:mean} is equivalent to $\hat{x}^{(i)}(k)$ and therefore the distributed
position solution delivered by Algorithm~1 is equivalent to the centralized one of~\eqref{eq:centralprobextended}. Since the same arguments hold for the control inputs and velocities the claim is proven. \hfill $\Box$

\begin{definition}
The vector $\left.x^{(i)}(k+1)\right|_{n_{i}}$ is the constructed local solution~\eqref{eq:localsolution} using an enlarged neighborhood size $n_i$ in the local problem $\Delta \mathbf{P}_i$~\eqref{eq:distributedprob}.
\end{definition}
\begin{definition}
The vector $\left.x(k+1)\right|_{\mathbf{n}}$ is the global solution of Algorithm~1 at step $k+1$, with $\mathbf{n} = (n_1, \dots, n_N)$. 
\end{definition}
Using the above definitions, we can prove the following theorem about the effect of an increased neighborhood size on the resulting algebraic connectivity. 
 
\begin{theorem}
If $\mathbf{n}_{1} \geq \mathbf{n}_{2}$ element-wise, then the algebraic connectivity of $\Delta L\left(\left.x(k+1)\right|_{\mathbf{n}_2}\right)$ is greater than equal to the one of $\Delta L\left(\left.x(k+1)\right|_{\mathbf{n}_1}\right)$, implying $\Delta L\left(\left.x(k+1)\right|_{\mathbf{n}_2}\right) \succeq \Delta L\left(\left.x(k+1)\right|_{\mathbf{n}_1}\right)$.
\label{theo:nton1}
\end{theorem}
\emph{Proof.}  By optimality and due to the linearity of $L$ on $x$ (Eq.~\eqref{eq.todo}), for each $i$ we can state
\footnotesize\vskip-.5cm
\begin{equation*}
  \Delta L\left(\left.\delta x^{(i)}(k+1)\right|_{\mathbf{n}_{2,i}}\right) \succeq \Delta L\left(\left.\delta x^{(i)}(k+1)\right|_{\mathbf{n}_{1,i}}\right)
\end{equation*}\normalsize
Multiplying by $\alpha_i$ and summing over $i$ leads to
\footnotesize\vskip-0.5cm
\begin{multline*}
  \sum_{i=1}^N \alpha_i \Delta L\left(\left.\delta x^{(i)}(k+1)\right|_{\mathbf{n}_{2,i}}\right) \succeq\\ \sum_{i=1}^N \alpha_i \Delta L\left(\left.\delta x^{(i)}(k+1)\right|_{\mathbf{n}_{1,i}}\right)
\end{multline*}\normalsize
By Lemma~\ref{lemma.deltax} the claim follows.
\hfill $\Box$

We note that Theorem~\ref{theo:nton1} holds considering one step horizon, i.e., from $k$ to $k+1$. Due to the non-linear/non-convex nature of the original problem~\eqref{eq:Kimprob}, this result does not hold in general from $k$ to $k+2$ or beyond, as we will show in the simulation experiments of Section~\ref{sec:results}.

Theorem~\ref{theo:nton1} is instrumental to construct a measure that can be used to decide locally on-line whether to increase or decrease the size $n_i$ of the enlarged neighborhood. This measure can be used to adapt $n_i$ to influence the trade-off between the increase of the algebraic connectivity or the reduction of the communication cost. For this purpose, we define two local relative sub-optimality measures with respect to a larger enlarged neighborhood size as
\begin{equation*}
  e_{i}^+ = 1 - \frac{\lambda_2(\Delta L_{i, n_i+1}(\left. x^{(i)}(k+1)\right|_{n_{i}}))}{\lambda_2(\Delta L_{i, n_i+1}(\left. x^{(i)}(k+1)\right|_{{n_i}+1}))}
\end{equation*}
\begin{equation*}
  e_{i}^- = 1 - \frac{\lambda_2(\Delta L_{i, n_i}(\left. x^{(i)}(k+1)\right|_{{n_i} -1}))}{\lambda_2(\Delta L_{i, {n_i}}(\left. x^{(i)}(k+1)\right|_{{n_{i}}}))}
\end{equation*}
which determine the sub-optimality of the local solutions~\eqref{eq:localsolution} with $n_i+1$ and $n_i-1$ with respect to the one obtained with $n_i$. In particular, $e_{i}^+$ measures the gain, in terms of local algebraic connectivity, one would have by increasing the enlarged neighborhood size from $n_i$ to $n_i + 1$, while $e_{i}^-$ measures the loss of local algebraic connectivity going from $n_i$ to $n_i - 1$. (We note that both $e_{i}^+$ and $e_{i}^-$ are non-negative due to Theorem~\ref{theo:nton1}). 

Given specific lower/upper thresholds for $e_{i}^+$ and $e_{i}^-$ the agents can decide locally to increase or decrease $n_i$ at the successive time step $k$, trading off larger communication efforts (for larger $n_i$) to smaller local algebraic connectivity increases (for smaller $n_i$), making Algorithm~1 adaptive. We note that although these sub-optimality measures are local, changing $n_i$ locally by each agent has an effect on the global solution as illustrated by the relation~\eqref{eq:delta} in Lemma~\ref{lemma.deltax}. We note also that in order to compute $e_{i}^+$ and $e_{i}^-$ it is necessary to solve three optimization problems of the kind~\eqref{eq:distributedprob} for each $i$. Since this can be computationally expensive, the agents can decide to determine $e_{i}^+$ and $e_{i}^-$ only once in a given number of discrete time steps.

\section{Numerical Results}\label{sec:results}

\begin{figure}
\centering
\psfrag{x}{\hskip-0.25cm \footnotesize $[1\,\, 0] u_i$} \psfrag{y}{\hskip-0.25cm \footnotesize $[0\,\, 1] u_i$}
    \includegraphics[width=0.20\textwidth]{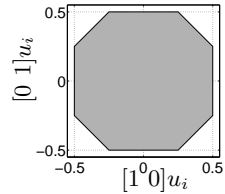}
    \caption{Polytopic constraint for $u_i$. The shaded region represents the set $\bar{\mathcal{U}}_i \subset \mathbb{R}^2$.}
    \label{fig:polytopic}
\end{figure}

In this section, we present numerical simulation results to
illustrate how the proposed distributed algorithm performs with respect to the
centralized scheme. We use a benchmark problem motivated by
\cite{Kim2006}. Our scenario considers $N = 10$ agents moving on a 2D
plane initially placed close to the horizontal axis and forming a
connected graph. The initial position vector is $x_i(0) =
\left[-6.75+1.5(i-1),\,y_i\right]^\top$, where $y_i$ is drawn from a Gaussian
distribution, with mean $0$ and standard deviation
$\sigma = 0.1$. Randomness is added to test the algorithm's
sensitivity to different initial conditions (due to the sequential convex programming approach). We consider the triples $(A_{1i}, A_{2i}, \bv_{1i})$ to be all equal to $(I_2
T_s/2, 0.75 I_2, I_2 T_s/2)$ with $T_s = 1$ (modeling a discrete-time double integrator dynamics), while all the $u_i$'s
are constrained in the polytopic region of
Figure~\ref{fig:polytopic}.
The other simulation parameters include the weighting function of
Figure~\ref{tab:poly}, $\rho_1 = 0.75$, $\rho_2 = 3$ and final time
$T = 300$. We performed and analyzed a total of $50$ simulation runs.

\begin{figure}
  % Requires \usepackage{graphicx}
  \psfrag{p1}{\hskip-0.2cm\tiny $\sqrt{\rho_1}$}
  \psfrag{p2}{\tiny  $\sqrt{\rho_2}$}
  \centering
  \includegraphics[trim = 0cm 0cm 6.5cm 0cm, clip=on, width=0.49\textwidth]{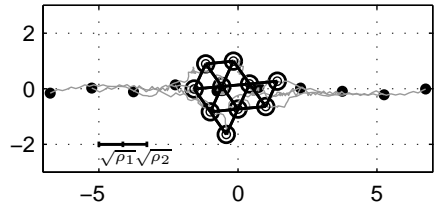}\\
  \caption{Centralized solution: the initial
positions are marked with black dots. The final positions
are marked with circles. The bold lines represent the final
communication graph and the thin lines the agent
trajectories.}\label{fig:Centralized}
\end{figure}

\begin{figure*}
  % Requires \usepackage{graphicx}
  \psfrag{d}{\hskip-2.9cm\small  $n_i = 1$, $k = 300$}
  \psfrag{f}{\hskip-2.9cm\small  $n_i = 2$, $k = 300$}
  \psfrag{h}{\hskip-2.9cm\small  $n_i = 3$, $k = 300$}
  \psfrag{l}{\hskip-2.9cm\small  Adaptive, $n_i(0) = 2$, $k = 300$}
  \psfrag{p1}{\hskip-0.2cm\tiny $\sqrt{\rho_1}$}
  \psfrag{p2}{\tiny  $\sqrt{\rho_2}$}
  \centering
  \includegraphics[trim = 0cm 1cm 0cm 1cm, clip=on, width=0.85\textwidth]{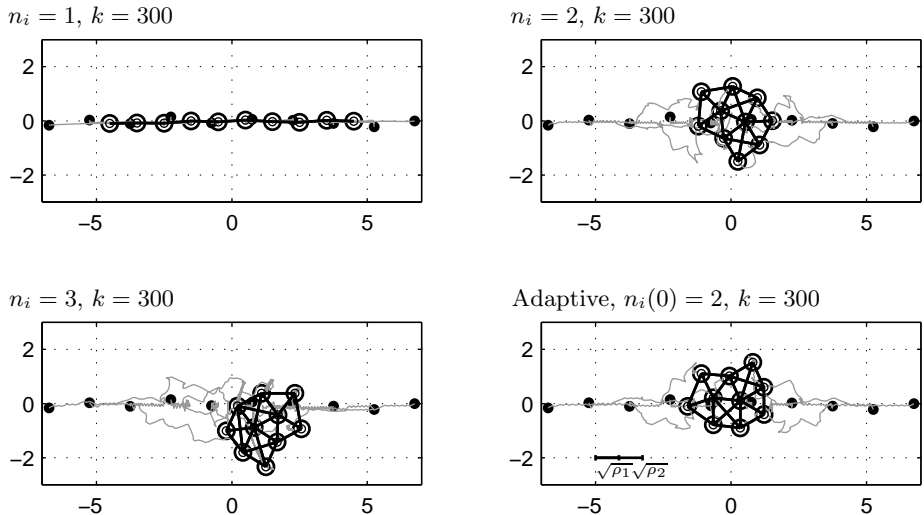}\\
  \caption{Simulation results of the distributed approach for various local neighborhood sizes $n_i$ (same $\forall i$ except in the adaptive case). The initial positions are marked with black dots. The final positions are marked with circles. The bold lines represent the final communication graph and the thin lines the agent trajectories. In the adaptive case, we start with $n_i=2$ $\forall i$ and at every $5$ discrete time step $k$ we compute $e_i^{+}$ and $e_i^{-}$. If $e_i^{+} > 0.05$ we increase $n_i$, if $e_i^{+} < 0.05$ and $e_i^- < 0.01$ we decrease $n_i$. }\label{fig:Distributed}
\end{figure*}

In Figures~\ref{fig:Centralized}-\ref{fig:Distributed}, an example of the
trajectories using the centralized and the distributed solutions are depicted. In the adaptive case, we start with $n_i=2$ for all agents and at every $5$ discrete time step $k$ we compute the sub-optimality measures. If the gain in increasing the enlarged neighborhood size is high enough, i.e., $e_i^{+} > 0.05$, we increase $n_i$, while if this gain is not high enough, i.e., $e_i^{+} < 0.05$, and the losses in decreasing the neighborhood size are not too big, i.e., $e_i^- < 0.01$, we decrease $n_i$ to reduce the communication and computation costs.

\begin{figure}
  % Requires \usepackage{graphicx}
  \centering
  \psfrag{a}{\hskip-2.5cm\small Discrete time instant $k$ in log scale}
  \psfrag{b}{\hskip-2.1cm\small Algebraic Connectivity $\lambda_2$}
  \psfrag{c}{\tiny Centralized}
  \psfrag{d}{\tiny Distributed $n_i = 1$}
  \psfrag{e}{\tiny Distributed $n_i = 2$}
  \psfrag{f}{\tiny Distributed $n_i = 3$}
  \psfrag{g}{\tiny Adaptive distr. $n_i(0) = 2$}
  \includegraphics[trim = 0cm 0cm 0cm 0cm, clip=on, width=0.48\textwidth]{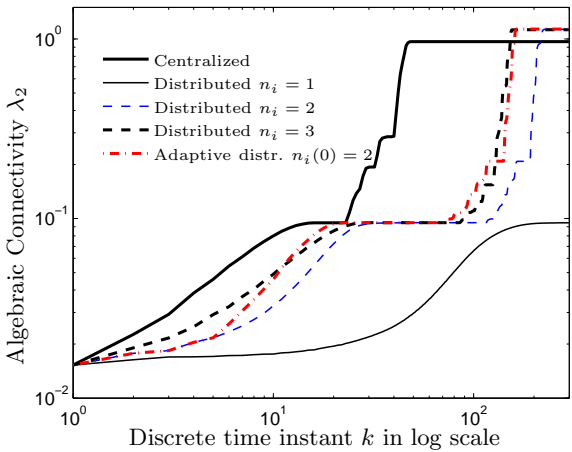}\\
 \caption{Algebraic connectivity as a function of time $k$ for both
the centralized and the distributed solutions.
}\label{fig:AlgebraicConnectivity}
\end{figure}

Figure~\ref{fig:AlgebraicConnectivity} shows, in the same
simulation, the algebraic connectivity as a function of the sampling
time $k$, and clearly illustrates the nonlinear/non-convex nature of
the problem. In fact, in this case, although the distributed
approximations are slower to converge than the centralized solution,
in the end they achieve a slightly better final $\lambda_2$.

Table~\ref{tab:comparison} shows the ratio between the final
connectivity of the distributed solution and the centralized one in
the $50$ simulation runs. For better comparison, we report that in the adaptive case $n_i = 2.2$ on average, with a maximum of $n_i = 5$.
\begin{table}     % Give a unique label
\small
\caption{Ratio between the final connectivity of the distributed solution and the centralized one. The adaptive case is indicated with $n_i(0)$. The cases $N = \{20, 40\}$ correspond to a random feasible initial configuration (not necessarily a line).}
  \begin{tabular}{p{0.1\textwidth}|p{0.020\textwidth}p{0.020\textwidth}p{0.020\textwidth}p{0.020\textwidth}||p{0.060\textwidth}|p{0.060\textwidth} }
  \hline
   & \multicolumn{4}{c||}{$N = 10$}& {$N = 20$}& {$N = 40$}\\
   Ratio $\displaystyle \frac{\lambda^{\mathrm{distr}}_2}{\lambda^{\mathrm{centr}}_2}$ & \rotatebox{-90}{\hskip-0.2cm$n_i = 1$} & \rotatebox{-90}{\hskip-0.2cm$n_i = 2$} & \rotatebox{-90}{\hskip-0.2cm$n_i = 3$} & \rotatebox{-90}{\hskip-0.2cm$n_i(0) = 2$~} & \hskip0.4cm\rotatebox{-90}{\hskip-0.2cm$n_i(0) = 2$~} & \hskip0.4cm\rotatebox{-90}{\hskip-0.2cm$n_i(0) = 2$~} \\ \hline
   $(0.1-0.3]$ & $50$ & $26$ & $0$ & $0$ & \hskip0.4cm$0$ & \hskip0.4cm$0$\\
   $(0.3-0.8]$ & $0$ & $0$ & $5$ & $4$ & \hskip0.4cm$3$& \hskip0.4cm$3$\\
   $(0.8-1.0]$ & $0$ & $12$ & $22$ & $21$ & \hskip0.4cm$21$& \hskip0.4cm$24$\\
   $(1.0-1.1]$ & $0$ & $12$ & $23$ & $25$ & \hskip0.4cm$26$& \hskip0.4cm$23$\\
  \hline
  \end{tabular}
\label{tab:comparison}  
\end{table}
We can observe that different choices of the local neighborhood sizes $n_i$ affect the final achieved $\lambda_2$. In particular, for the choice $n_i = 1$, the agents perform significantly worse than for other $n_i$. Furthermore, using the adaptive case, the final $\lambda_2$ is comparable with the centralized solution in most of the simulations (or even better). This is an important point, since the adaptive case use an enlarged neighborhood size of $n_i = 2.2$ (on average) and still obtains performances close or better than the fixed choice $n_i = 3$.

To further assess the proposed distributed algorithm, we include in Table~\ref{tab:comparison} simulation results for $N=\{20, 40\}$ robots starting from a feasible random configuration (not necessarily on a line) and using the adaptive algorithm with $n_i(0) = 2$. Each of these cases has been run $50$ times. We can observe that both in the $N = 20$ case (where the average $n_i$ is $2.7$) and in the $N = 40$ case (average $n_i = 2.6$), the results are in line with the conclusions that could be drawn for the case of $N = 10$. From these results one could conjecture both the scalability of Algorithm~\ref{algDSPD} (for the adaptive case) and its increased performances dealing with large systems. In particular, while the number of agents passes from $N=10$ to $N=40$, the averaged size of the enlarged neighborhood stays rather the same (and also the performance in term of final $\lambda_2)$. This means that the computational and communication efforts for the single agent stay the same. Thus, the gain of the distributed solution with respect of the centralized solution, in terms of computations and communications, increases.

\section{Conclusions}\label{sec:future}

We have presented a distributed solution to the maximization of the algebraic connectivity of the communication graph in a robotic network. Our characterization can handle more generic LTI agent dynamics than the methods available in the literature and the resulting optimization problem is proven to be feasible at each time step under reasonable assumptions. Furthermore the solution can be adjusted based on available resources using local relative sub-optimality measures to aid in adapting the neighborhood size to the agents' needs.

Simulation results confirm the efficacy of our distributed approach and show its practical applicability. Some open issues still remain and will be the focus of our future research. In particular, robustness of the proposed algorithm against estimation errors is currently being investigated. The applicability of the distributed scheme in a broader class of problem formulations involving LMI constraints is part of our research plans, as well as experimental validations.

%\begin{ack}                               % Place acknowledgements

%end{ack}

\bibliographystyle{plain}        % Include this if you use bibtex
\bibliography{PaperCollection2}           % and a bib file to produce the
                                 % bibliography (preferred). The
                                 % correct style is generated by
                                 % Elsevier at the time of printing.

%\appendix
%\renewcommand\thesection{Appendix \Alph{section}}

\appendix
%\renewcommand\thesection{Appendix \Alph{section}}

% !TEX root = Automatica2011_TReport.tex
\numberwithin{equation}{section}
\section{Generalization to LTI systems} \label{appendix1}

This appendix considers the issues related to the use of more general systems than~\eqref{eq:dynsyst} in the centralized problem~\eqref{eq:centralprobextended}. We start from a generalization of~\eqref{eq:dynsyst} considering the $(M+2)$-order system:
\footnotesize\vskip-.5cm
\begin{equation*}
  \left(\hskip-.05cm \begin{array}{c}
                x_i(\tau+1) \\
                v_i(\tau+1) \\
                y_{1i}(\tau+1) \\
                \vdots   \\
                y_{Mi}(\tau+1)
              \end{array}
            \hskip-.05cm\right)
   \hskip-.15cm = \hskip-.15cm\left( \begin{array}{ccccc}
                I_{3}& \star & \star & \cdots  \\
                0_3 & \star & \star & \cdots  \\
                0_3 & \star & \star & \cdots  \\
                \vdots & \vdots & \vdots & \ddots &  \\
              \end{array}
            \right)\hskip-.20cm \left(\hskip-.05cm \begin{array}{c}
                x_i(\tau) \\
                v_i(\tau) \\
                y_{1i}(\tau) \\
                \vdots   \\
                y_{Mi}(\tau)
              \end{array}\hskip-.05cm
            \right) \hskip-.05cm+\hskip-.05cm \left(\hskip-.05cm \begin{array}{c}
                0_{3}  \\
                \vdots\\
                \bv_{1i} I_3 \\
              \end{array}\hskip-.05cm
            \right) \hskip-.075cm u_i(\tau)
\end{equation*}\normalsize
where the stars represent non-zero elements and $\bv_{1i}\in
\mathbb{R}_0$. It is not difficult to see that Algorithm~1
is also applicable to these types of systems, under quite general
assumptions and minor modifications. The key idea is to compute the control actions every $M+2$ steps while the crucial drawback is that the larger $M+2$ is, the more $\rho_1$ has to be shrunk to accommodate the collision avoidance requirement (see condition~\eqref{eq:condition} which has to be generalized in this case in a straightforward manner).

Consider now the generic LTI system
\begin{equation*}
  \xs_i(\tau+1) = A_i \xs_i(\tau) + B_i u_i(\tau)
\end{equation*}
where the couple $(A_i, B_i)$ is controllable and where the state
can be partitioned as $(x_i(\tau)^\top, \xi_i(\tau)^\top)^\top$. In order to apply Algorithm~1, we need to characterize a modification of
the set $\mathcal{F}_i$ which is defined as:
\footnotesize\vskip-.75cm
\begin{multline*}
\xs_i(\tau) \in \mathcal{F}^T_i \Rightarrow \exists \{u_i(\tau),\dots,
u_i(\tau+T-1)\}\in \bar{\mathcal{U}}_i\,\, \mathrm{such~}\:\:
\mathrm{that}\\ A_i^T \xs_i(\tau) + \sum_{h = 0}^{T-1}
A_i^{T-1-h} B_i u(\tau+h) = (\x_i(\tau)^\top, 0)^\top , \, \forall \tau \in
\mathbb{N}_+
\end{multline*}
\normalsize
By computing $\mathcal{F}^T_i$, we can extend Algorithm~1 also to general
LTI systems, calculating the control every $T$ time steps.
However, several issues have to be addressed: \emph{(i)} the parameter $T$ is agent-dependent and it depends on $\tau$, making the determination of a single $\mathcal{F}^T_i$ quite complex; \emph{(ii)} the set $\mathcal{F}^T_i$ depends also on the position $x_i(\tau)$ restricting the area in which the agents can move; \emph{(iii)} since $T$ can be in general quite large, the condition on $\rho_1$ could be rather limiting and it could conflict with the requirements on $\mathcal{F}^T_i$.

\section{Proof of Theorem~\ref{theo:Q2}} \label{appendix.c}

At optimality the local constraints for the subproblem $\Delta \mathbf{P}_{p}$ in~\eqref{eq:distributedprob} are the following:
\footnotesize\vskip-0.75cm
\begin{subequations}\label{eq:pconstr}
\begin{eqnarray}
%\begin{array}{ll}
& \hskip-5cm \forall p| p \in \mathcal{J}_i^*\cap \mathcal{J}_j^*: \nonumber \\
\hat{\mathcal{Q}}_{2.1}:& \Delta f_d(x_i(k+1), x_j(k+1)) = \nonumber \\ & d^2_{ij}(k) + \b[ij]^\top (\delta \hat{x}_{pi}(k+1) - \delta
\hat{x}_{pj}(k+1)) > \hat{\rho}_{1ij}, \nonumber \\ & \label{eq:dummy2}\\ & \nonumber\\
 & \hskip-5.5cm \forall p| p \in \mathcal{J}_i^*: \nonumber \\
\hat{\mathcal{Q}}_{2.2}:& \hat{\xs}_{pi}(k+1) \in \hat{\mathcal{F}}_i = \mathcal{F}_i\\
\hat{\mathcal{Q}}_{2.3}:& \hat{\us}_{pi}(k) \in \hat{\mathcal{U}}_i \\
\hat{\mathcal{Q}}_{2.4}:& \hat{\xs}_{pi}(k+1) = \hat{\mathcal{D}}_i(\xs_i(k), \hat{\us}_{pi}(k))\qquad \label{eq:dummy32}
%\end{array} \\
\end{eqnarray}
\end{subequations}
\normalsize
The theorem claims that using the specified choice for $\mathcal{S}_{\Delta\hat{\mathcal{Q}}_{2i}}$, if we combine the local optimal solutions $(\hat{\xs}_{pi}(k+1), \hat{\us}_{pi}(k))$ which satisfy the local constraints~\eqref{eq:pconstr}, using the positive linear combinations~\eqref{eq:mean}, \eqref{eq:umean}, and \eqref{eq:vmean} we will obtain a couple $({\xs}(k+1), {\us}(k))$ that satisfies the constraint ${\Delta \mathcal{Q}}_{2}$ of the global problem~\eqref{eq:centralprobextended}. This is what we need to prove.

Consider $\hat{\mathcal{Q}}_{2.1}$ in~\eqref{eq:dummy2} and the positive linear combination for $x(k+1)$ in~\eqref{eq:mean}. By Lemma~\ref{lemma.deltaax} follows:
\footnotesize\vskip-0.75cm
\begin{multline}
d^2_{ij}(k) + \b[ij]^\top  (\delta {x}_{i}(k+1) - \delta
{x}_{j}(k+1)) =\\= d^2_{ij}(k) +  \sum_{p \in \mathcal{J}_i^*\cap
\mathcal{J}_j^*} \b[ij]^\top \alpha_p (\delta \hat{x}_{pi}(k+1) - \delta
\hat{x}_{pj}(k+1)) > \\ (1 - \bar{\alpha}_{ij}) d^2_{ij}(k) + \bar{\alpha}_{ij} \hat{\rho}_{1ij}
\label{eq:dummy4}
\end{multline}
\normalsize
For $x(k+1)$ it is required the satisfaction of the global constraint:
\begin{equation}
d^2_{ij}(k) + \b[ij]^\top (\delta {x}_{i}(k+1) - \delta
{x}_{j}(k+1)) > {\rho}_{1}
\label{eq:dummy3}
\end{equation}
which can be accomplished by selecting $\hat{\rho}_{1ij}$ such that:
\begin{equation}
(1 - \bar{\alpha}_{ij}) d^2_{ij}(k) + \bar{\alpha}_{ij} \hat{\rho}_{1ij} = {\rho}_{1}
\label{eq:dummy5}
\end{equation}
This gives the formula for $\hat{\rho}_{1ij}$ in~\eqref{eq:rhotilde}.

Consider the constraints $\hat{\mathcal{Q}}_{2.4}$ in~\eqref{eq:dummy32} on the agents' dynamics. For the positive linear combination~\eqref{eq:mean} the combined system dynamics becomes
\footnotesize \vskip-0.75cm
\begin{multline}
  \hskip-.45cm\left( \begin{array}{c}
                x_i(k+1) \\
                \displaystyle\sum_{p \in  \mathcal{J}_i^*} \alpha_p \hat{v}_{pi}(k+1) \\
              \end{array}
            \right)
   \hskip-.1cm=\hskip-.1cm \left( \begin{array}{cc}
                I_3 & \bar{\alpha}_i\hat{A}_{1i}(I_3 + \hat{A}_{2i}) \\
                0_3 & \bar{\alpha}_i \hat{A}^2_{2i} \\
              \end{array}
            \right) \hskip-.1cm \left( \begin{array}{c}
                x_i(k) \\
                v_i(k) \\
              \end{array}
            \right) +\\ \left( \begin{array}{cc}
                \hat{\bv}_{1i}\hat{A}_{1i} &  0_3 \\
                \hat{\bv}_{1i}\hat{A}_{2i} & \hat{\bv}_{1i} I_3 \\
              \end{array}
            \right) \sum_{p \in  \mathcal{J}_p^*} \alpha_p \hat{\us}_{pi}(k)
  \label{eq:update}
\end{multline}
\normalsize
Since the agents have to move according to the dynamical system~\eqref{eq:dynsystk} encoded in the global constraint
$\mathcal{Q}_{2.4}$ of~\eqref{eq:centralprobextended}, the update~\eqref{eq:update} and the state equation~\eqref{eq:dynsystk} have to be the same. It is not difficult to see that this is ensured by the choice $\hat{A}_{1i} = \bar{\alpha}^{-1}_i A_{1i}$, $\hat{A}_{2i} = A_{2i}$, $\hat{\bv}_{1i} = \bar{\alpha}_i \bv_{1i}$, and the linear combinations~\eqref{eq:umean} and~\eqref{eq:vmean} for the local control inputs $\hat{\us}_{pi}(k)$ and local velocities $\hat{v}_{pi}(k+1)$.

From the linear combination on the control~\eqref{eq:umean} and the global constraint ${\mathcal{Q}}_{2.3}$ in~\eqref{eq:centralprobextended} follows the specification for the local constraint $\hat{\mathcal{Q}}_{2.3}$ in~\eqref{eq:distributedprob}:
\begin{equation}
  \hat{\mathcal{U}}_i = \{\hat{\us}_{pi}(k) \in\mathbb{R}^3| H_i \hat{\us}_{pi} \leq \bar{\alpha}_i^{-1}\,h_i\}
\label{eq:utilde}
\end{equation}
from which $(\hat{H}_i, \hat{h}_i) = (H_i, \bar{\alpha}_i^{-1} h_i)$. We recall that the positive linear combination on the control input~\eqref{eq:umean} has been constructed in a way to steer the system~\eqref{eq:dynsystk} from the position $x(k)$ to the updated position $x(k+1)$ in~\eqref{eq:mean} while respecting the global constraints $\mathcal{Q}_{2.3}$ in~\eqref{eq:centralprobextended}.

Consider now $\hat{\mathcal{Q}}_{2.2}$ in~\eqref{eq:distributedprob}. We need to prove that if the local optimal states $\hat{\textbf{x}}_{pi}(k+1)$ belong to the set $\hat{\mathcal{F}}_i$ in~\eqref{eq:distributedprob}, then the updated state $\textbf{x}_i(k+1)$ constructed via the linear combinations on position~\eqref{eq:mean} and velocity~\eqref{eq:vmean} belongs to the set ${\mathcal{F}}_i$ as expressed in the global constraint $\mathcal{Q}_{2.2}$ in~\eqref{eq:centralprobextended}. First of all, it is straightforward to see that the local inequalities
\footnotesize\vskip-0.75cm
\begin{multline}
 -\left( \begin{array}{c}
                \hat{H}_i \hat{\bv}_{1i}^{-1} (I_3 + \hat{A}_{2i}) \\
                \hat{H}_i \hat{\bv}_{1i}^{-1}\hat{A}_{2i}(I_3 + 2 \hat{A}_{2i}) \\
              \end{array}
            \right) \hat{v}_{pi}(k+1)
   \leq \left( \begin{array}{c}
                \hat{h}_i \\
               \hat{h}_i \\
              \end{array}
            \right)
  \label{eq:localinequalities}
\end{multline}
\normalsize
are equivalent to the inequalities~\eqref{eq:F}, meaning that by construction $\hat{\mathcal{F}}_i = \mathcal{F}_i$. Recall that the set $\mathcal{F}_i$ does not constrain the position. Since the updated velocity $v_i(k+1)$ in~\eqref{eq:vmean} is obtained by a positive linear combination of local $\hat{v}_{pi}(k+1)$ then also $v_i(k+1)$ will satisfy the inequalities~\eqref{eq:localinequalities}, and therefore the updated state $\textbf{x}_i(k+1)$ belongs to $\mathcal{F}_i$.

Having ensured that with the choices of Theorem~\ref{theo:Q2} the positive linear combinations of the local solutions satisfy the constraints ${\mathcal{Q}}_{2.1} - {\mathcal{Q}}_{2.4}$ of~\eqref{eq:centralprobextended}, Theorem~\ref{theo:Q2} is proven. \hfill $\Box$

\end{document}